\documentclass[reprint,
    aps,prl,
    superscriptaddress,
    longbibliography,
    amsmath,amssymb,
    balancelastpage
]{revtex4-2}

\usepackage[utf8]{inputenc}
\usepackage[T1]{fontenc}

\usepackage{graphicx}   
\usepackage{dcolumn}    
\usepackage{bm}         

\usepackage{mathptmx}
\usepackage{etoolbox}
\usepackage{siunitx}
\usepackage{ORCIDinREVTeX}

\usepackage[unicode=true,
    colorlinks=true,
    linkcolor=blue,         
    citecolor=blue,         
    filecolor=blue,         
    urlcolor=blue]{hyperref}

\renewcommand*{\section}[1]{\paragraph{#1}}
\renewcommand*{\eqref}[1]{\hyperref[{#1}]{\textup{(\ref*{#1})}}}
\newcommand*{\figref}[1]{\hyperref[{#1}]{\textup{Fig.~\ref*{#1}}}}
\newcommand*{\secref}[1]{\hyperref[{#1}]{\textup{Sec.~\ref*{#1}}}}
\newcommand*{\tabref}[1]{\hyperref[{#1}]{\textup{Table~\ref*{#1}}}}
\newcommand*{\vect}[1]{\mathbf{#1}}  

\begin{document}
\title{Frequency-stable robust wireless power transfer based on high-order pseudo-Hermitian physics}

\author{Xianglin Hao}
\orcid{0000-0001-7149-0113}
\thanks{These authors contributed equally to this work.}

\author{Ke Yin}
\orcid{0000-0002-8534-216X}
\thanks{These authors contributed equally to this work.}

\author{Jianlong Zou}
\orcid{0000-0003-1489-0828}
\email{superzou@mail.xjtu.edu.cn}

\author{Ruibin Wang}

\author{Yuangen Huang}
\orcid{0000-0002-2486-2778}

\author{Xikui Ma}

\author{Tianyu Dong}
\orcid{0000-0003-4816-0073}
\email[Author to whom correspondence should be addressed. Please e-mail to: ]{tydong@mail.xjtu.edu.cn}
\affiliation{School of Electrical Engineering, Xi’an Jiaotong University, Xi’an 710049, China}%

\date{October 11, 2022}

\begin{abstract}
Non-radiative wireless power transfer (WPT) technology has made considerable progress with the application of the parity-time (PT) symmetry concept. In this letter, we extend the standard second-order PT-symmetric Hamiltonian to high-order symmetric tridiagonal pseudo-Hermitian Hamiltonian, relaxing the limitation of multi-source/multi-load system based on non-Hermitian physics. We proposed a three-mode pseudo-Hermitian dual-transmitter-single-receiver circuit and demonstrate that robust efficiency and stable frequency WPT can be achieved even though PT-symmetry is not satisfied as usual. In addition, no active tuning is required when the coupling coefficient between the intermediate transmitter and the receiver is changed. Moreover, the proposed system has an open frequency band gap with an abrupt frequency change at the phase transition point, which is expected to advance wireless sensing technologies.
\end{abstract}

\maketitle 

Wireless power transfer (WPT) shows exciting and promising applications in various fields where physical connections are not allowed. Driven by modern physical concepts \cite{krasnok2018coherently,song2021wireless,zeng2022efficient} and the ever-increasing practical demand \cite{lu2015wireless,dai2017safe,kan2018integrated,zhang2017wireless,li2020progress}, the \emph{so-called} magnetic resonance mechanism \cite{kurs2007wireless} for the non-radiative WPT technology has experienced rapid development in recent years. However, conventional WPT systems are not robust when source and load coils are dislocated or there exists a relative motion, yielding a variation of mutual coupling. Towards this end, with the introduction of parity-time (PT) symmetric systems \cite{bender1998real,schindler2011experimental,assawaworrarit2017robust,assawaworrarit2020robust,zhou2019nonlinear,wu2021position,hua2022pulse,yang2022observation}, the non-Hermitian theory \cite{ashida2020non,kawabata2019symmetry,el2018non} has been applied to WPT technology to address the long-standing robustness issue. With a real frequency spectrum, PT-symmetric WPT systems can self-select the operating frequency that corresponds to the maximum efficiency and hence guarantees optimal power transfer over a wide range of transfer distances without any active tunings \cite{assawaworrarit2017robust}. 

For a PT-symmetric WPT system, it requires gain elements covering the entire operating frequency range \cite{assawaworrarit2017robust,assawaworrarit2020robust,sakhdari2020robust}, which may increase the complexity and cost since the frequency variation range may be large when the natural resonant frequency of the coil is high. In addition, the PT-symmetric phase requires the coupling coefficient $k$ to be greater than the normalized gain/loss parameter $\gamma$ which is typically determined by the resistance of the gain/loss resonators \cite{schindler2012symmetric,chen2018generalized}; thus, the efficient working range of the system is usually limited by the resistance. Although high-order PT-symmetric systems are expected to achieve robust WPT with locked frequencies \cite{sakhdari2020robust,zeng2020high}, the coupling coefficients between adjacent resonators must be equal to support the PT-symmetric phase. As a consequence, it generally requires precise mechanical control of the system, which can lead to a substantial increase in complexity and cost.

For non-Hermitian systems, a real spectrum could exist not only in a PT-symmetric system but also in the \emph{so-called} pseudo-Hermitian one whose Hamiltonian satisfies $\bm{\mu} \bm{H} \bm{\mu}^{-1} = \bm{H}^{\dagger}$, where $\bm{\mu}$ is a Hermitian invertible operator \cite{mostafazadeh2002pseudo,mostafazadeh2002pseudoiii,grigoryan2022pseudo} and ``$\dagger$'' denotes the Hermitian conjugation, showing a more general class of Hamiltonians with real eigenvalues \cite{mostafazadeh2002pseudo,mostafazadeh2002pseudoiii}. In this letter, the conventional PT-symmetric WPT is extended to construct a multi-coil WPT system based on pseudo-Hermitian physics. As a prototype, we report a dual-transmitter-single-receiver system that incorporates nonlinear saturable gain elements into the source for WPT application. It is shown that, even though the system is not PT-symmetric, it still has real eigenfrequencies for varying coupling coefficients, and one of the eigenfrequencies does not change with respect to the coupling coefficient in the strong coupling region. Thus, we can achieve frequency-stable high-efficiency power transfer in the strong coupling region by utilizing the saturation characteristic of nonlinear gain elements; while the frequency only changes slightly in the weak coupling region.

\begin{figure}[!ht]
    \centering
    \includegraphics[width=3.0in]{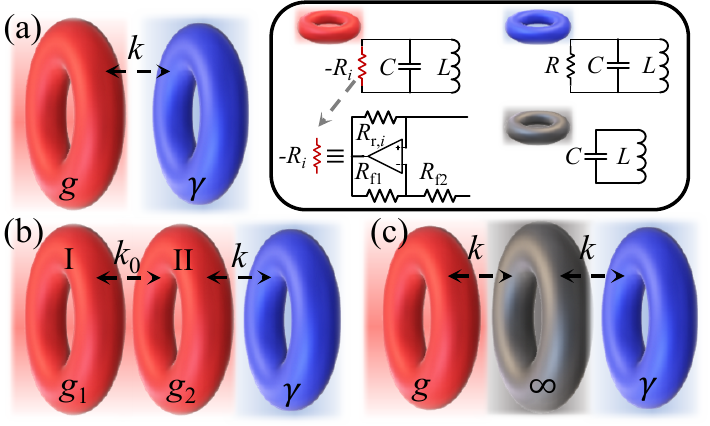}
    \caption{(color online) Schematics of (a) a standard PT-symmetric electronic dimer, (b) a three-mode pseudo-Hermitian system consisting of two gain (red) units and one loss (blue) unit, and (c) a conventional three-coil PT-symmetric system consisting of balanced gain and loss, as well as neutral (gray) electronic molecules. Here, $k$ or $k_0$ denotes the coupling coefficients between neighboring resonators; and the mutual coupling between the non-adjacent resonators [\emph{e.g.}, the transmitter I and receiver in (b)] is assumed to be negligible. }
    \label{fig:fig01}
\end{figure}

We start by constructing a class of higher-order WPT systems whose Hamiltonians are tridiagonal matrices from the standard second order PT-symmetric WPT system \cite{assawaworrarit2017robust,zhou2019nonlinear,wu2021position,hua2022pulse}, as shown in \figref{fig:fig01}(a). 
Within the coupled mode theory \cite{haus1991coupled}, the time evolution of the amplitudes of the transmitter and receiver resonator for the standard PT-symmetric dimer, denoted by $\vect{a} = [a_1,a_2]^\text{T}$, is governed by \cite{schindler2011experimental,schindler2012symmetric,assawaworrarit2017robust,assawaworrarit2020robust}
\begin{equation} \label{eq:system-equation}
    \text{i}\frac{\text{d}\vect{a}}{\text{d}t} =  \bm{H}_\text{std} \vect{a},
\end{equation}
where
\begin{equation} \label{eq:second-order-Hamiltonial}
     \bm{H}_\text{std} = 
     \begin{pmatrix}
        H_{11} & H_{12} \\
        H_{21} & H_{22} 
     \end{pmatrix} = \omega_0
     \begin{pmatrix}
        1+\text{i}\gamma & \kappa \\
        \kappa & 1-\text{i}\gamma 
     \end{pmatrix}
\end{equation}
is the Hamiltonian. Here, $H_{11} = H_{22}^* = \omega_0(1+\text{i}\gamma)$ with $\gamma$ denoting gain (or loss) and ``*'' denoting conjugation; and $H_{12}=H_{21} = \omega_0 \kappa$ with $\kappa$ denoting the mutual coupling between the resonators whose eigenfrequency are $\omega_0$. If the element $H_{11}$ or $H_{22}$ is replaced by a second-order block matrix of the same form as $\bm{H}_\text{std}$ \emph{per se}, and the mutual coupling is only considered for neighboring resonators, \emph{i.e.}, $H_{21} = H_{12}^\text{T} = [\kappa_{13}, \kappa_{23}]$ where $\kappa_{13} = 0$, we can form a three-mode systems, which can be a two-transmitter-one-receiver (see \figref{fig:fig01}(b)), a one-transmitter-two-receiver or a transmitter-repeater-receiver (see \figref{fig:fig01}(c)) system. In a similar manner, chain-like high-order systems can be constructed, whose Hamiltonian is symmetric tridiagonal if the mutual couplings for non-adjacent resonators are neglected. Since each resonator can be gainy, lossy or lossless, the corresponding Hamiltonian is a non-Hermitian symmetric matrix in general, whose diagonal elements can be complex and the off-diagonal elements are all real. For high-order systems, such a Hamiltonian corresponds to a multiple-transmitter multiple-receiver WPT system. In addition, a symmetric tridiagonal Hamiltonian can be PT-symmetric or pseudo-Hermitian, functioning the WPT applications. 

We focus on the Hamiltonian describing a dual-transmitter-single-receiver WPT system, as shown in \figref{fig:fig01}(b), \emph{i.e.}, 
\begin{equation} \label{eq:third-order-Hamiltonian}
     \bm{H}_\text{pse} = \frac{\omega_0}{2}
     \begin{pmatrix}
        2 + \text{i} g_1  & k_0     & 0 \\
        k_0   &  2 + \text{i} g_2  &  k \\
        0   &   k   & 2 - \text{i} \gamma
     \end{pmatrix},
\end{equation}
which has been proved to be a pseudo-Hermitian Hamiltonian under certain conditions \cite{xiong2021higher}. Here, $g_1$ and $g_2$ describe the strength of the gain in the transmitter resonators I and II, respectively; $\gamma$ is the loss constant of the receiver resonator; $k_0$ denotes the coupling coefficient between the two gain resonators; and $k$ denotes the coupling coefficient between the transmitter resonator II and the receiver resonator. The natural resonant frequencies of all the resonators are tuned to be $\omega_0$. Such a Hamiltonian \eqref{eq:third-order-Hamiltonian} can form a PT-symmetric system (see \figref{fig:fig01}(c)) when $g_1 = \gamma$, $g_2 = 0$ and $k_0 = k$ \cite{zeng2020high,sakhdari2020robust}. Moreover, even if it is not PT-symmetric, the Hamiltonian \eqref{eq:third-order-Hamiltonian} can still has real eigenvalues \cite{li2022high}. By solving the characteristic equation $\det\left(\omega \vect{I} - \bm{H}_\text{pse}\right) = 0$ (where $\vect{I}$ denotes an identity matrix), one arrives at
\begin{equation} \label{eq:ceqa}
\begin{split}
    \Delta \tilde{\omega} &\left[ \Delta \tilde{\omega}^2 + \frac{1}{4}( \gamma g_1+ \gamma g_2 - g_1 g_2 -k^2 -k_0^2) \right] + \\
    \frac{\text{i}}{2}  &\left[ (g_1+g_2-\gamma)\Delta \tilde{\omega}^2 +  \frac{1}{4} (g_1 g_2 \gamma +k_0^2 \gamma -k^2 g_1) \right] = 0,
\end{split}
\end{equation}
where $\Delta\tilde{\omega} = \tilde{\omega} - 1$ with $\tilde{\omega} = \omega/\omega_0$ being the normalized angular frequency. Instead of forcing the imaginary part of \eqref{eq:ceqa} to be zero \cite{assawaworrarit2017robust}, we can combine the like terms, which yields $\Delta\tilde{\omega} \left[ \Delta\tilde{\omega} \pm \sqrt{4\tilde{\kappa} -c^2}/4 - \text{i}c/4 \right] = 0$ where $\tilde{\kappa} = k_0^2 + k^2 + g_1 g_2 -\gamma (g_1 +g_2) $ and $c=g_1+g_2-\gamma$. Thus, in the strong coupling region when $k \geq \gamma$, three modes could exist, whose eigenfrequencies read as
\begin{subequations} \label{eq:eignw}
\begin{align}  
    \omega_1 &= \omega_0,  \\ 
    \omega_{2,3} &= \omega_0 \left( 1 \pm \sqrt{ 4\tilde{\kappa}-c^2 }/4 -\text{i}c/4 \right). 
\end{align}
\end{subequations}
According to \eqref{eq:ceqa}, provided that the criterion
\begin{equation}\label{eq:conds}
    (g_1 g_2 + k_0^2) \gamma = k^2 g_1
\end{equation}
is satisfied, the real mode $\omega = \omega_1$ could exist, no matter the modes $\omega_{2,3}$ are complex or real, which provides great design freedom. In the weak coupling region regime when $k < \gamma$, only two real-eigenvalue states exist and the corresponding eigenfrequencies read as
\begin{equation} \label{eq:ew23}
    \omega_{2,3} = \omega_0 \left( 1 \pm \sqrt{\tilde{\kappa}}/2 \right),
\end{equation}
and $\omega_1= \omega_0 - \text{i} c/2$ denoting the unstable states; while we have the criterion
\begin{equation}\label{eq:conds2}
    \left(g_1 g_2 \gamma +k_0^2 \gamma -k^2 g_1\right) c^{-1} = -\tilde{\kappa} 
\end{equation}
to support the stable-frequency mode for the weak coupling regime. We shall point out that, one can control the coupling regions by changing $\gamma$ and $k_0$ so that the phase transition may be independent of load, showing a fascinating realm in WPT applications. In a similar manner, one can handle the Hamiltonian \eqref{eq:third-order-Hamiltonian} of a single-transmitter dual-receiver system by the time-reversal transformation \cite{sachs1987physics,kawabata2019symmetry}. Furthermore, within the proposed method, one may be able to analyze the stable frequency mode of the pseudo-Hermitian extensions on anti-paritiy-time (APT) symmetry systems \cite{choi2018observation,luo2022quantum,kim2022visualization}  when the coupling coefficients are pure imaginary, and even systems with 
complex coupling coefficients \cite{grigoryan2022pseudo, arwas2022anyonic}.

We first consider the case when the strong coupling region supports three real modes when $c=0$. To design a pseudo-Hermitian WPT system, with the known load denoted by $\gamma$, one can determine  the two gain coefficients $g_1$ and $g_2$ of the system according to the coupling coefficients $k$ and $k_0$, as implicated by \eqref{eq:conds} or \eqref{eq:conds2}. In the strong coupling region when $k \geq \gamma$, the steady-state gains $g_1$ and $g_2$ can be obtained by solving $c=0$ and \eqref{eq:conds} as $g_{1,2} = \left[ \gamma^2 \mp k^2 \pm \sqrt{(\gamma^2 - k^2)^2 +4\gamma^2 k_0^2} \right]/(2\gamma)$. It is evident that the sum of gain coefficients of the system is always equal to the loss coefficient $\gamma$, \emph{i.e.}, $g_1 + g_2 = \gamma$. When $k \gg k_0 = \gamma$, the gain coefficient $g_2$ dominates as $g_2 \gg g_1$. As $k$ is decreased, $g_2$ is decreased while $g_1$ is increased; when $g_1=\gamma$ and $g_2=0$, the system is in the PT-symmetric state. In the weak coupling region when $k<\gamma$, $g_1$ and $g_2$ are related according to \eqref{eq:conds2}. For a given value of $g_1$ or $g_2$, one can determine the other gain; for instance, when $g_2=0$ for the intermediate resonator II in the weak coupling region, one obtain the gain for the source resonator I reads as $g_1=\left[\gamma^2 + k_0^2 -\sqrt{(\gamma^2 + k_0^2)^2 -4\gamma^2 k^2}\right]/(2\gamma)$, which is rapidly decreased to zero as $k$ is decreased.

\begin{figure}[!ht]
    \centering
    \includegraphics[width=3.3in]{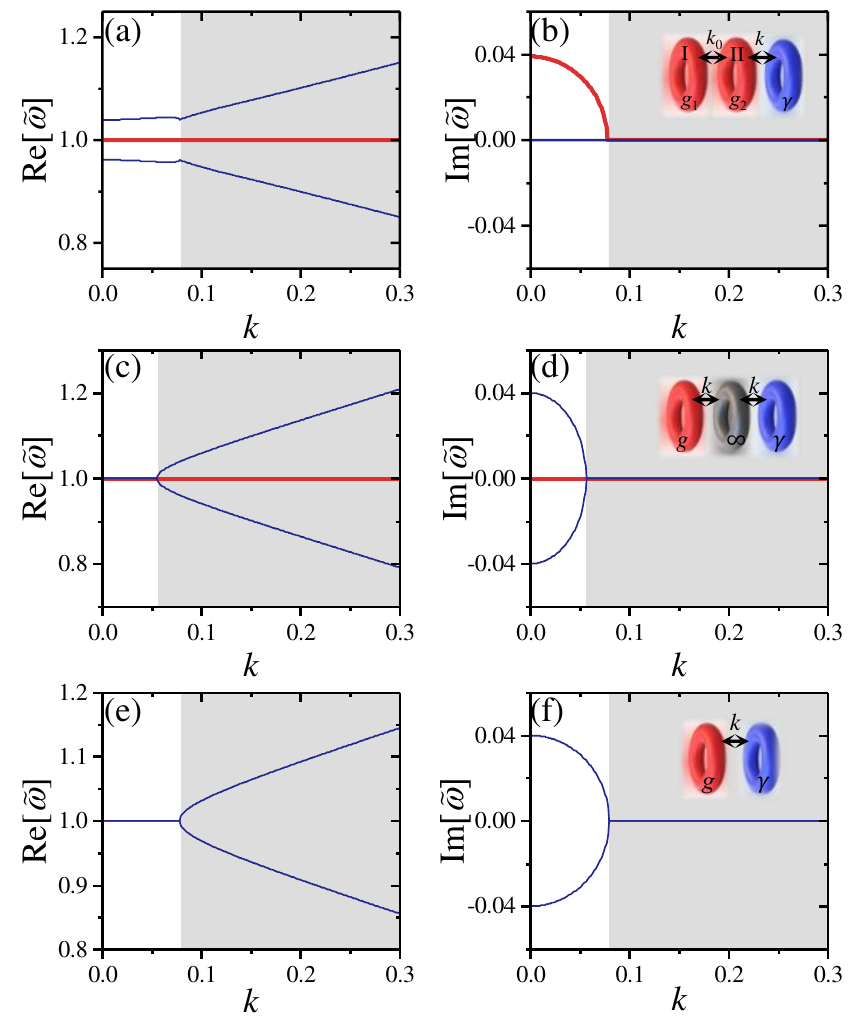}
    \caption{(color online) Evolution of the real (left panel) and imaginary (right panel) parts of the normalized eigenfrequencies $\tilde{\omega}= \omega/\omega_0$ as a function of the coupling coefficient $k$ for various non-Hermitian systems, \emph{i.e.}, (a,b) three-mode pseudo-Hermitian system, (c,d) three-coil PT-symmetric system, and (e,f) standard PT-symmetric system. For all the three systems, the loss parameter reads $\gamma = 0.078$. Here, the blue curves denotes the conjugate solution of $\omega$; while the red curves denotes the additional solution of $\omega_1$ for three-mode systems. The ideal three-coil PT-symmetric WPT system always works in the state [red line in (c,d)] where the frequency does not change with respect to the coupling coefficient $k$. The strong coupling regions are denoted in shaded.}
    \label{fig:fig02}
\end{figure}

In this letter, we focus on the scenario when $k_0=\gamma$ to achieve the same phase transition point as the standard PT-symmetric system for the comparison. As shown in \figref{fig:fig02}(a–b), the proposed three-mode pseudo-Hermitian system has a unique frequency characteristic with an open band gap, while the conventional PT-symmetric systems have an exceptional point in the transition from the strong coupling region to the weak coupling region [see in \figref{fig:fig02}(c-f)]. Such a unique frequency characteristic reflects the asymmetry of the pseudo-Hermitian system, which can be tuned by changing the ratio of coupling coefficients, \emph{i.e.}, $k_0/k$. In addition, the system has a steady-state frequency that does not vary with $k$ in the strong coupling region when $k \geq \gamma$, while its steady-state frequency can be designed in a narrow frequency range in the weak coupling region when $k<\gamma$. Therefore, such a three-coil pseudo-Hermitian system can be used to realize WPT with almost stable operating frequency, even in the weak coupling regime. Note that the steady-state frequency will inevitably undergo discontinuous jumps when the system whose operating frequency is set to $\omega_1$ in the strong coupling region enters into the weak coupling region through the phase transition point, showing a possibility to further enhance the sensitivity in wireless sensing applications \cite{chen2018generalized,yin2022wireless,yang2021ultrarobust}. 

As shown in \figref{fig:fig02}(a), the eigenfrequency $\omega_1 = \omega_0$ is locked to the natural frequency of the $LC$ tank, which is independent of the coupling coefficient $k$ in the strong coupling region. For the eigen-states denoted by $\omega_{2,3}$, only a small derivation of $\pm \sqrt{\tilde{\kappa}}/2$ of the eigenfrequency is presented, both for the strong and weak coupling regime. Therefore, high-efficiency power transfer always appear around $\omega_0$. Moreover, the pseudo-Hermitian system relaxes the limitation that the coupling coefficients $k_0$ and $k$ must be identical for the conventional three-coil PT-symmetric system. Compared with the standard second-order PT-symmetric system, whose eigenfrequencies are sensitive to the change of $k$ and $\gamma$ [see \figref{fig:fig02}(e) and \figref{fig:fig02}(f)] in the strong coupling region, the proposed pseudo-Hermitian system is more friendly when designing the gain elements, which is expected to achieve a higher power transmission by combining the two transmitters.

\begin{figure}[!ht]
    \centering
    \includegraphics[width=3.3in]{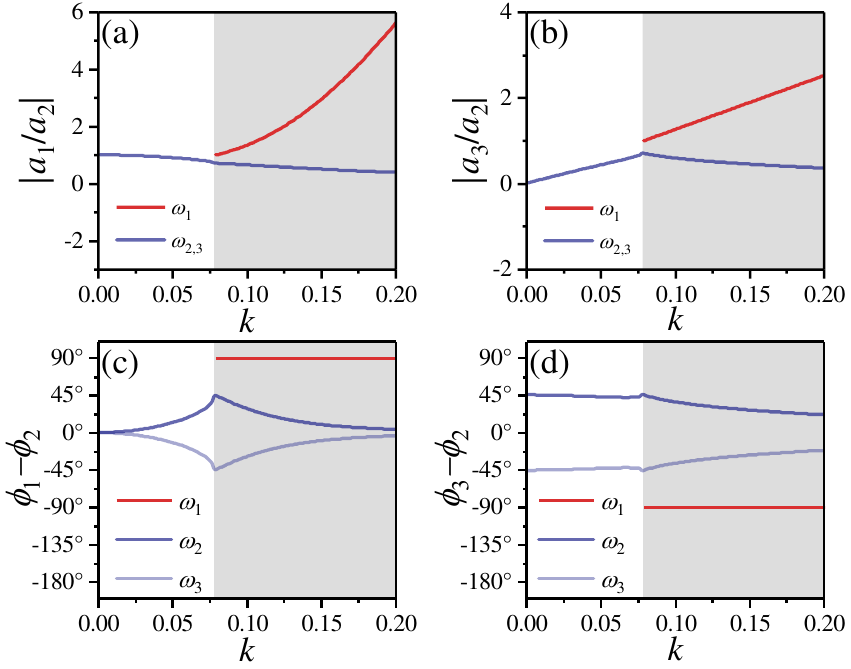}
    \caption{(color online) Evolution of the amplitude ratios (a)  $|a_1/a_2|$ and (b) $|a_3/a_2|$, and the phase differences $\phi_1-\phi_2$ and (d) $\phi_3-\phi_2$ as functions of the coupling coefficient $k$ at the steady state. In all the sub-figures, the vertical grey reference line indicates $k=\gamma=0.078$.}
    \label{fig:fig03}
\end{figure}

The proposed three-mode pseudo-Hermitian system has unique power transfer properties. In the strong coupling region when $k \geq \gamma$, the ratios of the resonators' amplitudes respectively reads $a_1/a_2 = \text{i} k_0/g_1$ and $a_3/a_2 = -\text{i} k/\gamma$ for the mode $\omega_1 = \omega_0$; while in the weak coupling region when $k < \gamma$, they are $a_1/a_2=k_0/\left(-\text{i} g_1 \pm \sqrt{\tilde{\kappa} +\gamma g_2}\right)$ and $a_3/a_2 = k/\left(\text{i} \gamma \pm \sqrt{\tilde{\kappa} + \gamma g_2}\right)$, respectively. Figures \ref{fig:fig03}(a) and \ref{fig:fig03}(b) plot the amplitude ratios as functions of the coupling coefficients. When the steady-state frequency switches from $\omega_1$ ($k \geq \gamma$ in the strong coupling region) to $\omega_{2,3}$ ($k<\gamma$ in the weak coupling region) at the phase transition point $k_0=0.078$, a small abrupt change of the amplitude ratio appears. Although it would lead to a small abrupt efficiency change around the transition point, it does not affect the stability of the system. In the strong coupling region, the range of the amplitude ratio for the steady-state denoted by $\omega_1$ is greater than that for the state denoted by $\omega_2$ or $\omega_3$. Such a varying amplitude ratios demonstrates the possibility of implementing a transformer in WPT systems, which may be beneficial for the applications of extracting energy from high-voltage power transmission lines. Moreover, as illustrated in \figref{fig:fig03}(c) and \figref{fig:fig03}(d), when the system frequency is locked at $\omega_1 = \omega_0$ in the strong coupling region, the phase differences between the resonators are constant, which is more friendly for monitoring the system operation state.

The proposed system delivers the combined power from the two transmitters to the load. As a result, the transmission efficiency of the proposed three-coil pseudo-Hermitian system can be calculated by dividing the load power by the total power. For the proposed system, the varying amplitude ratio with respect to the coupling coefficient $k$ implicates that the transfer efficiency may not be a constant. Nevertheless, provided that the quality factors of the resonators are large enough, the transmission efficiency of the three-coil system is basically stable in the strong coupling region \cite{sakhdari2020robust,seo2019comparative}. In fact, coils used in WPT systems are usually lossy. For resonant coils with finite quality factor, it is possible to evaluate the transmission efficiency to determine whether a pseudo-Hermitian WPT system is practical. Suppose the losses of transmitter I, transmitter II and the receiver are denoted by $\gamma_\text{s1}$, $\gamma_\text{s2}$ and $\gamma_\text{s3}$, respectively and the loss of the load on the receiving side reads $\gamma_\ell=\gamma - \gamma_\text{s3}$, the steady-state power transmission efficiency $\eta_\text{PTE}=\gamma_\ell |a_1|^2/(\gamma_\text{s1} |a_1|^2+\gamma_\text{s2} |a_2|^2+\gamma |a_3|^2)$, as a function of coupling coefficient $k$, can be expressed as
\begin{equation} \label{eq:tf}
    \eta_\text{PTE} = \left\{
    \begin{array}{lr}
        \frac{\gamma_\ell}{\gamma} \left( 1+\frac{\gamma_\text{s1} \gamma k_0^2}{g_1^2 k^2} +\frac{\gamma_\text{s2} \gamma}{k^2} \right)^{-1},  &  k \geq \gamma, \\[1.0em]
        \frac{\gamma_\ell}{\gamma}  \left( 1+\alpha \frac{\gamma_\text{s1}}{\gamma} + \beta \frac{\gamma_\text{s2}}{\gamma} \right)^{-1},  & k < \gamma,
    \end{array} \right.
\end{equation}
where $\alpha = [k^2(k^2+k_0^2+g_1^2-g_1 \gamma)]/[k_0^2 (k^2+k_0^2+\gamma^2-g_1 \gamma)]$ and $\beta = (k^2+k_0^2+\gamma^2-g_1 \gamma)/k^2$. In the strong coupling region, the larger the coupling coefficient $k$ is, the lower the efficiency $\eta_\text{PTE}$ will be. However, for the WPT system operating at several megahertz, since the loss constant $\gamma$ of resonant coils is typically very small (about on the magnitude of $10^{-3}$), the decay rate of the transmission efficiency is relatively small when the coupling coefficient $k$ is increased, which is acceptable. In addition, the peak transmission efficiency is still close to unity in the limit $\gamma_{\text{s}n} \ll \gamma_\ell$ where $n=1,2,3$. 

The implementation of gain elements in non-Hermitian systems is crucial for WPT applications. According to the criterion \eqref{eq:conds}, the gain strengths $g$'s are related to the coupling coefficient $k$. To avoid active tuning of gain when $k$ changes, nonlinear gain elements consisting of an operational amplifier (op-amp) can be adopted \cite{schindler2011experimental,assawaworrarit2017robust,yin2022wireless} , as illustrated in the inset of \figref{fig:fig01}. Due to the nonlinearity and saturation behaviors of op-amps \cite{assawaworrarit2017robust}, the actual gain $g_n(|a_n|)$ ($n=1,2$) of the transmitters depends on the mode amplitude $|a_n|$, which will decrease as $|a_n|$ is increased beyond the threshold. If the op-amps are configured to saturation with the initial unsaturated gains $g_{n,i}$ being respectively greater than or equal to the corresponding steady-state desired gains, \emph{i.e.}, $g_{n,i} \geq g_n$, the system will eventually go into the stable oscillating state \cite{supplementary}.
\begin{figure}[!ht]
    \centering
    \includegraphics[width=3.0in]{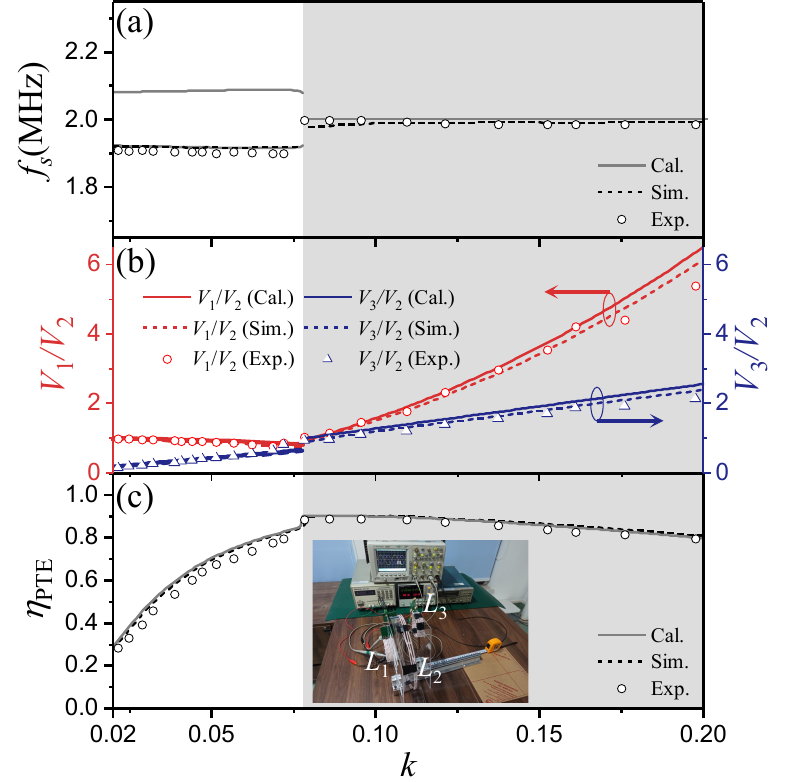}
    \caption{(color online) (a) Steady-state frequency $f_s$, (b) voltage ratios $V_1/V_2$ (red for left y-axis) and $V_3/V_2$ (blue for right $y$-axis), and (c) power transmission efficiency $\eta_\text{PTE}$ as functions of the coupling coefficient $k$. Theoretical (Cal.), simulated (Sim.) and experimental (Exp.) results are illustrated by solid lines, dashed lines and hollow markers, respectively. The shading regions indicate $k \geq \gamma = 0.078$. The inset of (c) shows a photo of our experiment setup.}
    \label{fig:fig04}
\end{figure}

We have verified the theory via circuit simulations and experiments of the two-transmitter-one-receiver system \cite{supplementary}. All the coils are tuned to be resonant at 2~\si{MHz}, and the initial gains are $g_{1i}=0.113$ and $g_{2i}=0.005$, respectively. As a result, only a single steady-state $\omega_1=\omega_0$ exhibits in the strong coupling region since $\omega_{2,3}$ are complex. \figref{fig:fig04} shows the steady-state operating frequency $f_s$, resonators’ voltage ratios and transmission efficiency $\eta_\text{PTE}$ with respect to the coupling coefficient $k$. The experimental results show a good agreement with simulated and theoretical results. As shown in \figref{fig:fig04}(a), the operating frequency is keeping about 2~\si{MHz} as the coupling coefficient $k$ varies from 0.078 to approximately 0.2 in the strong coupling region, without any tuning of the circuit. In the strong coupling region, the system has almost a stable transmission efficiency, as illustrated in \figref{fig:fig04}(b). In addition, the efficiency would slightly drop due to the over-coupling as $k$ is increased. Both the simulation and experimental results demonstrate the transition of the steady-state frequency around $k=\gamma$, which implicates that the frequency band gap exhibits in the system. Since the amplitude ratios of the resonator are discontinuous at the transition point $k=\gamma$, a slight abrupt change of the transmission efficiency appears. Such a transition would only introduce slight changes of the frequency and efficiency in the weak coupling region, which would not affect the stable operation of the WPT system. 

By optimizing the loss constant $\gamma$ and/or the coupling coefficient $k_0$, one may further improve the efficiency and the robustness of the proposed system. Moreover, when the relay resonator is source-free and the loss is considered, \emph{i.e.}, $g_2= -\gamma_\text{s2}$, we can also achieve a frequency-stable three-coil WPT system provided that the gain of the transmitter reads $g_1 = k_0^2 \gamma / (k^2 + \gamma_{s2}\gamma)$. Given that the amplitude ratios of the system are fixed, highly efficient voltage-adjustable switch-mode amplifiers could be designed for a practical system to achieve constant power output. In addition, based on the concept of generalized PT symmetry \cite{chen2018generalized,ye2020tunability}, the inductance of the proposed system can be further optimized to facilitate coil design. For practical systems, high-efficiency switch-mode amplifiers \cite{assawaworrarit2020robust,zhou2019nonlinear,wu2021position} can be used as gain elements to achieve a high overall system efficiency.

In summary, we have reported a three-mode pseudo-Hermitian system constituting of nonlinear gain-saturating elements for robust dynamic wireless power transfer operated at almost fixed frequency without any active tuning. The proposed system always has real eigenfrequencies even though it is not PT-symmetric. The system also exhibits varying amplitude ratios, promising multi-functional platforms for transformers and wireless power transfer. Our work reveals that the multi-mode electronic circuit concept based on non-Hermitian physics with nonlinear gain holds promise for efficient dynamic wireless power transfer without active tuning. Meanwhile, our theory is expected to be applicable to handle multi-coil WPT systems when relay coils are lossy, while the existing PT-symmetric systems generally fail \cite{kim2022wide,qu2022wireless}. Although the dual-transmitter-single-receiver topology is considered here, one may readily apply the time-reversal transformation \cite{kawabata2019symmetry} to handle the single-transmitter-dual-receiver circuit \cite{mohseni2022one}. Moreover, our research may open a door to the investigation on the multiple-transmitter-one-receiver, one-transmitter-multiple-receiver, and multiple-transmitter-multiple-receiver WPT systems by virtue of the pseudo-Hermitian physics. It has also been demonstrated that band gaps can appear in the frequency band of three-mode non-Hermitian systems by integrating nonlinear gains, which is expected to advance wireless sensing technologies.

The authors acknowledge support from the National Natural Science Foundation of China (NSFC) with Grant No. 51977165.

\bibliography{main} 

\providecommand{\noopsort}[1]{}\providecommand{\singleletter}[1]{#1}%
\begin{thebibliography}{43}%
\makeatletter
\providecommand \@ifxundefined [1]{%
 \@ifx{#1\undefined}
}%
\providecommand \@ifnum [1]{%
 \ifnum #1\expandafter \@firstoftwo
 \else \expandafter \@secondoftwo
 \fi
}%
\providecommand \@ifx [1]{%
 \ifx #1\expandafter \@firstoftwo
 \else \expandafter \@secondoftwo
 \fi
}%
\providecommand \natexlab [1]{#1}%
\providecommand \enquote  [1]{``#1''}%
\providecommand \bibnamefont  [1]{#1}%
\providecommand \bibfnamefont [1]{#1}%
\providecommand \citenamefont [1]{#1}%
\providecommand \href@noop [0]{\@secondoftwo}%
\providecommand \href [0]{\begingroup \@sanitize@url \@href}%
\providecommand \@href[1]{\@@startlink{#1}\@@href}%
\providecommand \@@href[1]{\endgroup#1\@@endlink}%
\providecommand \@sanitize@url [0]{\catcode `\\12\catcode `\$12\catcode
  `\&12\catcode `\#12\catcode `\^12\catcode `\_12\catcode `\%12\relax}%
\providecommand \@@startlink[1]{}%
\providecommand \@@endlink[0]{}%
\providecommand \url  [0]{\begingroup\@sanitize@url \@url }%
\providecommand \@url [1]{\endgroup\@href {#1}{\urlprefix }}%
\providecommand \urlprefix  [0]{URL }%
\providecommand \Eprint [0]{\href }%
\providecommand \doibase [0]{https://doi.org/}%
\providecommand \selectlanguage [0]{\@gobble}%
\providecommand \bibinfo  [0]{\@secondoftwo}%
\providecommand \bibfield  [0]{\@secondoftwo}%
\providecommand \translation [1]{[#1]}%
\providecommand \BibitemOpen [0]{}%
\providecommand \bibitemStop [0]{}%
\providecommand \bibitemNoStop [0]{.\EOS\space}%
\providecommand \EOS [0]{\spacefactor3000\relax}%
\providecommand \BibitemShut  [1]{\csname bibitem#1\endcsname}%
\let\auto@bib@innerbib\@empty
\bibitem [{\citenamefont {Krasnok}\ \emph {et~al.}(2018)\citenamefont
  {Krasnok}, \citenamefont {Baranov}, \citenamefont {Generalov}, \citenamefont
  {Li},\ and\ \citenamefont {Al{\`u}}}]{krasnok2018coherently}%
  \BibitemOpen
  \bibfield  {author} {\bibinfo {author} {\bibfnamefont {A.}~\bibnamefont
  {Krasnok}}, \bibinfo {author} {\bibfnamefont {D.~G.}\ \bibnamefont
  {Baranov}}, \bibinfo {author} {\bibfnamefont {A.}~\bibnamefont {Generalov}},
  \bibinfo {author} {\bibfnamefont {S.}~\bibnamefont {Li}},\ and\ \bibinfo
  {author} {\bibfnamefont {A.}~\bibnamefont {Al{\`u}}},\ }\bibfield  {title}
  {\bibinfo {title} {Coherently enhanced wireless power transfer},\ }\href
  {https://doi.org/10.1103/PhysRevLett.120.143901} {\bibfield  {journal}
  {\bibinfo  {journal} {Phys. Rev. Lett.}\ }\textbf {\bibinfo {volume} {120}},\
  \bibinfo {pages} {143901} (\bibinfo {year} {2018})}\BibitemShut {NoStop}%
\bibitem [{\citenamefont {Song}\ \emph {et~al.}(2021)\citenamefont {Song},
  \citenamefont {Jayathurathnage}, \citenamefont {Zanganeh}, \citenamefont
  {Krasikova}, \citenamefont {Smirnov}, \citenamefont {Belov}, \citenamefont
  {Kapitanova}, \citenamefont {Simovski}, \citenamefont {Tretyakov},\ and\
  \citenamefont {Krasnok}}]{song2021wireless}%
  \BibitemOpen
  \bibfield  {author} {\bibinfo {author} {\bibfnamefont {M.}~\bibnamefont
  {Song}}, \bibinfo {author} {\bibfnamefont {P.}~\bibnamefont
  {Jayathurathnage}}, \bibinfo {author} {\bibfnamefont {E.}~\bibnamefont
  {Zanganeh}}, \bibinfo {author} {\bibfnamefont {M.}~\bibnamefont {Krasikova}},
  \bibinfo {author} {\bibfnamefont {P.}~\bibnamefont {Smirnov}}, \bibinfo
  {author} {\bibfnamefont {P.}~\bibnamefont {Belov}}, \bibinfo {author}
  {\bibfnamefont {P.}~\bibnamefont {Kapitanova}}, \bibinfo {author}
  {\bibfnamefont {C.}~\bibnamefont {Simovski}}, \bibinfo {author}
  {\bibfnamefont {S.}~\bibnamefont {Tretyakov}},\ and\ \bibinfo {author}
  {\bibfnamefont {A.}~\bibnamefont {Krasnok}},\ }\bibfield  {title} {\bibinfo
  {title} {Wireless power transfer based on novel physical concepts},\ }\href
  {https://doi.org/10.1038/s41928-021-00658-x} {\bibfield  {journal} {\bibinfo
  {journal} {Nat. Electron.}\ }\textbf {\bibinfo {volume} {4}},\ \bibinfo
  {pages} {707} (\bibinfo {year} {2021})}\BibitemShut {NoStop}%
\bibitem [{\citenamefont {Zeng}\ \emph {et~al.}(2022)\citenamefont {Zeng},
  \citenamefont {Guo}, \citenamefont {Zhu}, \citenamefont {Fan}, \citenamefont
  {Li}, \citenamefont {Jiang}, \citenamefont {Li}, \citenamefont {Jiang},
  \citenamefont {Yang}, \citenamefont {Sun},\ and\ \citenamefont
  {Chen}}]{zeng2022efficient}%
  \BibitemOpen
  \bibfield  {author} {\bibinfo {author} {\bibfnamefont {C.}~\bibnamefont
  {Zeng}}, \bibinfo {author} {\bibfnamefont {Z.}~\bibnamefont {Guo}}, \bibinfo
  {author} {\bibfnamefont {K.}~\bibnamefont {Zhu}}, \bibinfo {author}
  {\bibfnamefont {C.}~\bibnamefont {Fan}}, \bibinfo {author} {\bibfnamefont
  {G.}~\bibnamefont {Li}}, \bibinfo {author} {\bibfnamefont {J.}~\bibnamefont
  {Jiang}}, \bibinfo {author} {\bibfnamefont {Y.}~\bibnamefont {Li}}, \bibinfo
  {author} {\bibfnamefont {H.}~\bibnamefont {Jiang}}, \bibinfo {author}
  {\bibfnamefont {Y.}~\bibnamefont {Yang}}, \bibinfo {author} {\bibfnamefont
  {Y.}~\bibnamefont {Sun}},\ and\ \bibinfo {author} {\bibfnamefont
  {H.}~\bibnamefont {Chen}},\ }\bibfield  {title} {\bibinfo {title} {Efficient
  and stable wireless power transfer based on the non-{Hermitian} physics},\
  }\href {https://doi.org/10.1088/1674-1056/ac3815} {\bibfield  {journal}
  {\bibinfo  {journal} {Chin. Phys. B}\ }\textbf {\bibinfo {volume} {31}},\
  \bibinfo {pages} {010307} (\bibinfo {year} {2022})}\BibitemShut {NoStop}%
\bibitem [{\citenamefont {Lu}\ \emph {et~al.}(2015)\citenamefont {Lu},
  \citenamefont {Niyato}, \citenamefont {Wang}, \citenamefont {Kim},\ and\
  \citenamefont {Han}}]{lu2015wireless}%
  \BibitemOpen
  \bibfield  {author} {\bibinfo {author} {\bibfnamefont {X.}~\bibnamefont
  {Lu}}, \bibinfo {author} {\bibfnamefont {D.}~\bibnamefont {Niyato}}, \bibinfo
  {author} {\bibfnamefont {P.}~\bibnamefont {Wang}}, \bibinfo {author}
  {\bibfnamefont {D.~I.}\ \bibnamefont {Kim}},\ and\ \bibinfo {author}
  {\bibfnamefont {Z.}~\bibnamefont {Han}},\ }\bibfield  {title} {\bibinfo
  {title} {Wireless charger networking for mobile devices: {Fundamentals},
  standards, and applications},\ }\href
  {https://doi.org/10.1109/MWC.2015.7096295} {\bibfield  {journal} {\bibinfo
  {journal} {IEEE Wirel. Commun.}\ }\textbf {\bibinfo {volume} {22}},\ \bibinfo
  {pages} {126} (\bibinfo {year} {2015})}\BibitemShut {NoStop}%
\bibitem [{\citenamefont {Dai}\ \emph {et~al.}(2017)\citenamefont {Dai},
  \citenamefont {Liu}, \citenamefont {Chen}, \citenamefont {Wu}, \citenamefont
  {He}, \citenamefont {Liu},\ and\ \citenamefont {Ma}}]{dai2017safe}%
  \BibitemOpen
  \bibfield  {author} {\bibinfo {author} {\bibfnamefont {H.}~\bibnamefont
  {Dai}}, \bibinfo {author} {\bibfnamefont {Y.}~\bibnamefont {Liu}}, \bibinfo
  {author} {\bibfnamefont {G.}~\bibnamefont {Chen}}, \bibinfo {author}
  {\bibfnamefont {X.}~\bibnamefont {Wu}}, \bibinfo {author} {\bibfnamefont
  {T.}~\bibnamefont {He}}, \bibinfo {author} {\bibfnamefont {A.~X.}\
  \bibnamefont {Liu}},\ and\ \bibinfo {author} {\bibfnamefont {H.}~\bibnamefont
  {Ma}},\ }\bibfield  {title} {\bibinfo {title} {Safe charging for wireless
  power transfer},\ }\href {https://doi.org/10.1109/INFOCOM.2014.6848041}
  {\bibfield  {journal} {\bibinfo  {journal} {IEEE/ACM Trans. Netw.}\ }\textbf
  {\bibinfo {volume} {25}},\ \bibinfo {pages} {3531} (\bibinfo {year}
  {2017})}\BibitemShut {NoStop}%
\bibitem [{\citenamefont {Kan}\ \emph {et~al.}(2018)\citenamefont {Kan},
  \citenamefont {Lu}, \citenamefont {Nguyen}, \citenamefont {Mercier},\ and\
  \citenamefont {Mi}}]{kan2018integrated}%
  \BibitemOpen
  \bibfield  {author} {\bibinfo {author} {\bibfnamefont {T.}~\bibnamefont
  {Kan}}, \bibinfo {author} {\bibfnamefont {F.}~\bibnamefont {Lu}}, \bibinfo
  {author} {\bibfnamefont {T.-D.}\ \bibnamefont {Nguyen}}, \bibinfo {author}
  {\bibfnamefont {P.~P.}\ \bibnamefont {Mercier}},\ and\ \bibinfo {author}
  {\bibfnamefont {C.~C.}\ \bibnamefont {Mi}},\ }\bibfield  {title} {\bibinfo
  {title} {Integrated coil design for {EV} wireless charging systems using
  {LCC} compensation topology},\ }\href
  {https://doi.org/10.1109/TPEL.2018.2794448} {\bibfield  {journal} {\bibinfo
  {journal} {IEEE Trans. Power Electron.}\ }\textbf {\bibinfo {volume} {33}},\
  \bibinfo {pages} {9231} (\bibinfo {year} {2018})}\BibitemShut {NoStop}%
\bibitem [{\citenamefont {Zhang}\ \emph {et~al.}(2017)\citenamefont {Zhang},
  \citenamefont {Qian}, \citenamefont {Wu}, \citenamefont {Kong},\ and\
  \citenamefont {Lu}}]{zhang2017wireless}%
  \BibitemOpen
  \bibfield  {author} {\bibinfo {author} {\bibfnamefont {S.}~\bibnamefont
  {Zhang}}, \bibinfo {author} {\bibfnamefont {Z.}~\bibnamefont {Qian}},
  \bibinfo {author} {\bibfnamefont {J.}~\bibnamefont {Wu}}, \bibinfo {author}
  {\bibfnamefont {F.}~\bibnamefont {Kong}},\ and\ \bibinfo {author}
  {\bibfnamefont {S.}~\bibnamefont {Lu}},\ }\bibfield  {title} {\bibinfo
  {title} {Wireless charger placement and power allocation for maximizing
  charging quality},\ }\href {https://doi.org/10.1109/TMC.2017.2771425}
  {\bibfield  {journal} {\bibinfo  {journal} {IEEE Trans. Mob. Comput.}\
  }\textbf {\bibinfo {volume} {17}},\ \bibinfo {pages} {1483} (\bibinfo {year}
  {2017})}\BibitemShut {NoStop}%
\bibitem [{\citenamefont {Li}\ \emph {et~al.}(2020)\citenamefont {Li},
  \citenamefont {Zhang}, \citenamefont {Song},\ and\ \citenamefont
  {Huang}}]{li2020progress}%
  \BibitemOpen
  \bibfield  {author} {\bibinfo {author} {\bibfnamefont {L.}~\bibnamefont
  {Li}}, \bibinfo {author} {\bibfnamefont {X.}~\bibnamefont {Zhang}}, \bibinfo
  {author} {\bibfnamefont {C.}~\bibnamefont {Song}},\ and\ \bibinfo {author}
  {\bibfnamefont {Y.}~\bibnamefont {Huang}},\ }\bibfield  {title} {\bibinfo
  {title} {Progress, challenges, and perspective on metasurfaces for ambient
  radio frequency energy harvesting},\ }\href
  {https://doi.org/10.1063/1.5140966} {\bibfield  {journal} {\bibinfo
  {journal} {Appl. Phys. Lett.}\ }\textbf {\bibinfo {volume} {116}},\ \bibinfo
  {pages} {060501} (\bibinfo {year} {2020})}\BibitemShut {NoStop}%
\bibitem [{\citenamefont {Kurs}\ \emph {et~al.}(2007)\citenamefont {Kurs},
  \citenamefont {Karalis}, \citenamefont {Moffatt}, \citenamefont
  {Joannopoulos}, \citenamefont {Fisher},\ and\ \citenamefont
  {Solja\v{c}i\`{c}}}]{kurs2007wireless}%
  \BibitemOpen
  \bibfield  {author} {\bibinfo {author} {\bibfnamefont {A.}~\bibnamefont
  {Kurs}}, \bibinfo {author} {\bibfnamefont {A.}~\bibnamefont {Karalis}},
  \bibinfo {author} {\bibfnamefont {R.}~\bibnamefont {Moffatt}}, \bibinfo
  {author} {\bibfnamefont {J.~D.}\ \bibnamefont {Joannopoulos}}, \bibinfo
  {author} {\bibfnamefont {P.}~\bibnamefont {Fisher}},\ and\ \bibinfo {author}
  {\bibfnamefont {M.}~\bibnamefont {Solja\v{c}i\`{c}}},\ }\bibfield  {title}
  {\bibinfo {title} {Wireless power transfer via strongly coupled magnetic
  resonances},\ }\href {https://doi.org/10.1126/science.1143254} {\bibfield
  {journal} {\bibinfo  {journal} {Science}\ }\textbf {\bibinfo {volume}
  {317}},\ \bibinfo {pages} {83} (\bibinfo {year} {2007})}\BibitemShut
  {NoStop}%
\bibitem [{\citenamefont {Bender}\ and\ \citenamefont
  {Boettcher}(1998)}]{bender1998real}%
  \BibitemOpen
  \bibfield  {author} {\bibinfo {author} {\bibfnamefont {C.~M.}\ \bibnamefont
  {Bender}}\ and\ \bibinfo {author} {\bibfnamefont {S.}~\bibnamefont
  {Boettcher}},\ }\bibfield  {title} {\bibinfo {title} {Real spectra in
  non-{Hermitian} {Hamiltonians} having {PT} symmetry},\ }\href
  {https://doi.org/10.1103/PhysRevLett.80.5243} {\bibfield  {journal} {\bibinfo
   {journal} {Phys. Rev. Lett.}\ }\textbf {\bibinfo {volume} {80}},\ \bibinfo
  {pages} {5243} (\bibinfo {year} {1998})}\BibitemShut {NoStop}%
\bibitem [{\citenamefont {Schindler}\ \emph {et~al.}(2011)\citenamefont
  {Schindler}, \citenamefont {Li}, \citenamefont {Zheng}, \citenamefont
  {Ellis},\ and\ \citenamefont {Kottos}}]{schindler2011experimental}%
  \BibitemOpen
  \bibfield  {author} {\bibinfo {author} {\bibfnamefont {J.}~\bibnamefont
  {Schindler}}, \bibinfo {author} {\bibfnamefont {A.}~\bibnamefont {Li}},
  \bibinfo {author} {\bibfnamefont {M.~C.}\ \bibnamefont {Zheng}}, \bibinfo
  {author} {\bibfnamefont {F.~M.}\ \bibnamefont {Ellis}},\ and\ \bibinfo
  {author} {\bibfnamefont {T.}~\bibnamefont {Kottos}},\ }\bibfield  {title}
  {\bibinfo {title} {Experimental study of active {LRC} circuits with {PT}
  symmetries},\ }\href {https://doi.org/10.1103/PhysRevA.84.040101} {\bibfield
  {journal} {\bibinfo  {journal} {Phys. Rev. A}\ }\textbf {\bibinfo {volume}
  {84}},\ \bibinfo {pages} {040101(R)} (\bibinfo {year} {2011})}\BibitemShut
  {NoStop}%
\bibitem [{\citenamefont {Assawaworrarit}\ \emph {et~al.}(2017)\citenamefont
  {Assawaworrarit}, \citenamefont {Yu},\ and\ \citenamefont
  {Fan}}]{assawaworrarit2017robust}%
  \BibitemOpen
  \bibfield  {author} {\bibinfo {author} {\bibfnamefont {S.}~\bibnamefont
  {Assawaworrarit}}, \bibinfo {author} {\bibfnamefont {X.}~\bibnamefont {Yu}},\
  and\ \bibinfo {author} {\bibfnamefont {S.}~\bibnamefont {Fan}},\ }\bibfield
  {title} {\bibinfo {title} {Robust wireless power transfer using a nonlinear
  parity-time-symmetric circuit},\ }\href {https://doi.org/10.1038/nature22404}
  {\bibfield  {journal} {\bibinfo  {journal} {Nature}\ }\textbf {\bibinfo
  {volume} {546}},\ \bibinfo {pages} {387} (\bibinfo {year}
  {2017})}\BibitemShut {NoStop}%
\bibitem [{\citenamefont {Assawaworrarit}\ and\ \citenamefont
  {Fan}(2020)}]{assawaworrarit2020robust}%
  \BibitemOpen
  \bibfield  {author} {\bibinfo {author} {\bibfnamefont {S.}~\bibnamefont
  {Assawaworrarit}}\ and\ \bibinfo {author} {\bibfnamefont {S.}~\bibnamefont
  {Fan}},\ }\bibfield  {title} {\bibinfo {title} {Robust and efficient wireless
  power transfer using a switch-mode implementation of a nonlinear parity-time
  symmetric circuit},\ }\href {https://doi.org/10.1038/s41928-020-0399-7}
  {\bibfield  {journal} {\bibinfo  {journal} {Nat. Electron.}\ }\textbf
  {\bibinfo {volume} {3}},\ \bibinfo {pages} {273} (\bibinfo {year}
  {2020})}\BibitemShut {NoStop}%
\bibitem [{\citenamefont {Zhou}\ \emph {et~al.}(2019)\citenamefont {Zhou},
  \citenamefont {Zhang}, \citenamefont {Xiao}, \citenamefont {Qiu},\ and\
  \citenamefont {Chen}}]{zhou2019nonlinear}%
  \BibitemOpen
  \bibfield  {author} {\bibinfo {author} {\bibfnamefont {J.}~\bibnamefont
  {Zhou}}, \bibinfo {author} {\bibfnamefont {B.}~\bibnamefont {Zhang}},
  \bibinfo {author} {\bibfnamefont {W.}~\bibnamefont {Xiao}}, \bibinfo {author}
  {\bibfnamefont {D.}~\bibnamefont {Qiu}},\ and\ \bibinfo {author}
  {\bibfnamefont {Y.}~\bibnamefont {Chen}},\ }\bibfield  {title} {\bibinfo
  {title} {Nonlinear parity-time-symmetric model for constant efficiency
  wireless power transfer: {Application} to a drone-in-flight wireless charging
  platform},\ }\href {https://doi.org/10.1109/TIE.2018.2864515} {\bibfield
  {journal} {\bibinfo  {journal} {IEEE Trans. Ind. Electron.}\ }\textbf
  {\bibinfo {volume} {66}},\ \bibinfo {pages} {4097} (\bibinfo {year}
  {2019})}\BibitemShut {NoStop}%
\bibitem [{\citenamefont {Wu}\ \emph {et~al.}(2022)\citenamefont {Wu},
  \citenamefont {Zhang},\ and\ \citenamefont {Jiang}}]{wu2021position}%
  \BibitemOpen
  \bibfield  {author} {\bibinfo {author} {\bibfnamefont {L.}~\bibnamefont
  {Wu}}, \bibinfo {author} {\bibfnamefont {B.}~\bibnamefont {Zhang}},\ and\
  \bibinfo {author} {\bibfnamefont {Y.}~\bibnamefont {Jiang}},\ }\bibfield
  {title} {\bibinfo {title} {Position-independent {CC/CV} wireless {EV}
  charging system without dual-side communication and {DC-DC} converter},\
  }\href {https://doi.org/10.1109/TIE.2021.3108702} {\bibfield  {journal}
  {\bibinfo  {journal} {IEEE Trans. Ind. Electron.}\ }\textbf {\bibinfo
  {volume} {69}},\ \bibinfo {pages} {7930} (\bibinfo {year}
  {2022})}\BibitemShut {NoStop}%
\bibitem [{\citenamefont {Hua}\ \emph {et~al.}(2022)\citenamefont {Hua},
  \citenamefont {Chau}, \citenamefont {Liu},\ and\ \citenamefont
  {Tian}}]{hua2022pulse}%
  \BibitemOpen
  \bibfield  {author} {\bibinfo {author} {\bibfnamefont {Z.}~\bibnamefont
  {Hua}}, \bibinfo {author} {\bibfnamefont {K.}~\bibnamefont {Chau}}, \bibinfo
  {author} {\bibfnamefont {W.}~\bibnamefont {Liu}},\ and\ \bibinfo {author}
  {\bibfnamefont {X.}~\bibnamefont {Tian}},\ }\bibfield  {title} {\bibinfo
  {title} {Pulse frequency modulation for parity-time-symmetric wireless power
  transfer system},\ }\href {https://doi.org/10.1109/TMAG.2022.3153499}
  {\bibfield  {journal} {\bibinfo  {journal} {IEEE Trans. Magn.}\ } (\bibinfo
  {year} {2022})}\BibitemShut {NoStop}%
\bibitem [{\citenamefont {Yang}\ \emph {et~al.}(2022)\citenamefont {Yang},
  \citenamefont {Li}, \citenamefont {Ding}, \citenamefont {Xu}, \citenamefont
  {Zhu},\ and\ \citenamefont {Zhu}}]{yang2022observation}%
  \BibitemOpen
  \bibfield  {author} {\bibinfo {author} {\bibfnamefont {X.}~\bibnamefont
  {Yang}}, \bibinfo {author} {\bibfnamefont {J.}~\bibnamefont {Li}}, \bibinfo
  {author} {\bibfnamefont {Y.}~\bibnamefont {Ding}}, \bibinfo {author}
  {\bibfnamefont {M.}~\bibnamefont {Xu}}, \bibinfo {author} {\bibfnamefont
  {X.-F.}\ \bibnamefont {Zhu}},\ and\ \bibinfo {author} {\bibfnamefont
  {J.}~\bibnamefont {Zhu}},\ }\bibfield  {title} {\bibinfo {title} {Observation
  of transient parity-time symmetry in electronic systems},\ }\href
  {https://doi.org/10.1103/PhysRevLett.128.065701} {\bibfield  {journal}
  {\bibinfo  {journal} {Phys. Rev. Lett.}\ }\textbf {\bibinfo {volume} {128}},\
  \bibinfo {pages} {065701} (\bibinfo {year} {2022})}\BibitemShut {NoStop}%
\bibitem [{\citenamefont {Ashida}\ \emph {et~al.}(2020)\citenamefont {Ashida},
  \citenamefont {Gong},\ and\ \citenamefont {Ueda}}]{ashida2020non}%
  \BibitemOpen
  \bibfield  {author} {\bibinfo {author} {\bibfnamefont {Y.}~\bibnamefont
  {Ashida}}, \bibinfo {author} {\bibfnamefont {Z.}~\bibnamefont {Gong}},\ and\
  \bibinfo {author} {\bibfnamefont {M.}~\bibnamefont {Ueda}},\ }\bibfield
  {title} {\bibinfo {title} {Non-{Hermitian} physics},\ }\href
  {https://doi.org/10.1080/00018732.2021.1876991} {\bibfield  {journal}
  {\bibinfo  {journal} {Adv. Phys.}\ }\textbf {\bibinfo {volume} {69}},\
  \bibinfo {pages} {249} (\bibinfo {year} {2020})}\BibitemShut {NoStop}%
\bibitem [{\citenamefont {Kawabata}\ \emph {et~al.}(2019)\citenamefont
  {Kawabata}, \citenamefont {Shiozaki}, \citenamefont {Ueda},\ and\
  \citenamefont {Sato}}]{kawabata2019symmetry}%
  \BibitemOpen
  \bibfield  {author} {\bibinfo {author} {\bibfnamefont {K.}~\bibnamefont
  {Kawabata}}, \bibinfo {author} {\bibfnamefont {K.}~\bibnamefont {Shiozaki}},
  \bibinfo {author} {\bibfnamefont {M.}~\bibnamefont {Ueda}},\ and\ \bibinfo
  {author} {\bibfnamefont {M.}~\bibnamefont {Sato}},\ }\bibfield  {title}
  {\bibinfo {title} {Symmetry and topology in non-{Hermitian} physics},\ }\href
  {https://doi.org/10.1103/PhysRevX.9.041015} {\bibfield  {journal} {\bibinfo
  {journal} {Phys. Rev. X}\ }\textbf {\bibinfo {volume} {9}},\ \bibinfo {pages}
  {041015} (\bibinfo {year} {2019})}\BibitemShut {NoStop}%
\bibitem [{\citenamefont {El-Ganainy}\ \emph {et~al.}(2018)\citenamefont
  {El-Ganainy}, \citenamefont {Makris}, \citenamefont {Khajavikhan},
  \citenamefont {Musslimani}, \citenamefont {Rotter},\ and\ \citenamefont
  {Christodoulides}}]{el2018non}%
  \BibitemOpen
  \bibfield  {author} {\bibinfo {author} {\bibfnamefont {R.}~\bibnamefont
  {El-Ganainy}}, \bibinfo {author} {\bibfnamefont {K.~G.}\ \bibnamefont
  {Makris}}, \bibinfo {author} {\bibfnamefont {M.}~\bibnamefont {Khajavikhan}},
  \bibinfo {author} {\bibfnamefont {Z.~H.}\ \bibnamefont {Musslimani}},
  \bibinfo {author} {\bibfnamefont {S.}~\bibnamefont {Rotter}},\ and\ \bibinfo
  {author} {\bibfnamefont {D.~N.}\ \bibnamefont {Christodoulides}},\ }\bibfield
   {title} {\bibinfo {title} {Non-{Hermitian} physics and {PT} symmetry},\
  }\href {https://doi.org/10.1038/nphys4323} {\bibfield  {journal} {\bibinfo
  {journal} {Nat. Phys.}\ }\textbf {\bibinfo {volume} {14}},\ \bibinfo {pages}
  {11} (\bibinfo {year} {2018})}\BibitemShut {NoStop}%
\bibitem [{\citenamefont {Sakhdari}\ \emph {et~al.}(2020)\citenamefont
  {Sakhdari}, \citenamefont {Hajizadegan},\ and\ \citenamefont
  {Chen}}]{sakhdari2020robust}%
  \BibitemOpen
  \bibfield  {author} {\bibinfo {author} {\bibfnamefont {M.}~\bibnamefont
  {Sakhdari}}, \bibinfo {author} {\bibfnamefont {M.}~\bibnamefont
  {Hajizadegan}},\ and\ \bibinfo {author} {\bibfnamefont {P.-Y.}\ \bibnamefont
  {Chen}},\ }\bibfield  {title} {\bibinfo {title} {Robust extended-range
  wireless power transfer using a higher-order {PT}-symmetric platform},\
  }\href {https://doi.org/10.1103/PhysRevResearch.2.013152} {\bibfield
  {journal} {\bibinfo  {journal} {Phys. Rev. Res.}\ }\textbf {\bibinfo {volume}
  {2}},\ \bibinfo {pages} {013152} (\bibinfo {year} {2020})}\BibitemShut
  {NoStop}%
\bibitem [{\citenamefont {Schindler}\ \emph {et~al.}(2012)\citenamefont
  {Schindler}, \citenamefont {Lin}, \citenamefont {Lee}, \citenamefont
  {Ramezani}, \citenamefont {Ellis},\ and\ \citenamefont
  {Kottos}}]{schindler2012symmetric}%
  \BibitemOpen
  \bibfield  {author} {\bibinfo {author} {\bibfnamefont {J.}~\bibnamefont
  {Schindler}}, \bibinfo {author} {\bibfnamefont {Z.}~\bibnamefont {Lin}},
  \bibinfo {author} {\bibfnamefont {J.}~\bibnamefont {Lee}}, \bibinfo {author}
  {\bibfnamefont {H.}~\bibnamefont {Ramezani}}, \bibinfo {author}
  {\bibfnamefont {F.~M.}\ \bibnamefont {Ellis}},\ and\ \bibinfo {author}
  {\bibfnamefont {T.}~\bibnamefont {Kottos}},\ }\bibfield  {title} {\bibinfo
  {title} {{PT}-symmetric electronics},\ }\href
  {https://doi.org/10.1088/1751-8113/45/44/444029} {\bibfield  {journal}
  {\bibinfo  {journal} {J. Phys. A: Math. Theor.}\ }\textbf {\bibinfo {volume}
  {45}},\ \bibinfo {pages} {444029} (\bibinfo {year} {2012})}\BibitemShut
  {NoStop}%
\bibitem [{\citenamefont {Chen}\ \emph {et~al.}(2018)\citenamefont {Chen},
  \citenamefont {Sakhdari}, \citenamefont {Hajizadegan}, \citenamefont {Cui},
  \citenamefont {Cheng}, \citenamefont {El-Ganainy},\ and\ \citenamefont
  {Al{\`u}}}]{chen2018generalized}%
  \BibitemOpen
  \bibfield  {author} {\bibinfo {author} {\bibfnamefont {P.-Y.}\ \bibnamefont
  {Chen}}, \bibinfo {author} {\bibfnamefont {M.}~\bibnamefont {Sakhdari}},
  \bibinfo {author} {\bibfnamefont {M.}~\bibnamefont {Hajizadegan}}, \bibinfo
  {author} {\bibfnamefont {Q.}~\bibnamefont {Cui}}, \bibinfo {author}
  {\bibfnamefont {M.~M.-C.}\ \bibnamefont {Cheng}}, \bibinfo {author}
  {\bibfnamefont {R.}~\bibnamefont {El-Ganainy}},\ and\ \bibinfo {author}
  {\bibfnamefont {A.}~\bibnamefont {Al{\`u}}},\ }\bibfield  {title} {\bibinfo
  {title} {Generalized parity--time symmetry condition for enhanced sensor
  telemetry},\ }\href {https://doi.org/10.1038/s41928-018-0072-6} {\bibfield
  {journal} {\bibinfo  {journal} {Nat. Electron.}\ }\textbf {\bibinfo {volume}
  {1}},\ \bibinfo {pages} {297} (\bibinfo {year} {2018})}\BibitemShut {NoStop}%
\bibitem [{\citenamefont {Zeng}\ \emph {et~al.}(2020)\citenamefont {Zeng},
  \citenamefont {Sun}, \citenamefont {Li}, \citenamefont {Li}, \citenamefont
  {Jiang}, \citenamefont {Yang},\ and\ \citenamefont {Chen}}]{zeng2020high}%
  \BibitemOpen
  \bibfield  {author} {\bibinfo {author} {\bibfnamefont {C.}~\bibnamefont
  {Zeng}}, \bibinfo {author} {\bibfnamefont {Y.}~\bibnamefont {Sun}}, \bibinfo
  {author} {\bibfnamefont {G.}~\bibnamefont {Li}}, \bibinfo {author}
  {\bibfnamefont {Y.}~\bibnamefont {Li}}, \bibinfo {author} {\bibfnamefont
  {H.}~\bibnamefont {Jiang}}, \bibinfo {author} {\bibfnamefont
  {Y.}~\bibnamefont {Yang}},\ and\ \bibinfo {author} {\bibfnamefont
  {H.}~\bibnamefont {Chen}},\ }\bibfield  {title} {\bibinfo {title} {High-order
  parity-time symmetric model for stable three-coil wireless power transfer},\
  }\href {https://doi.org/10.1103/PhysRevApplied.13.034054} {\bibfield
  {journal} {\bibinfo  {journal} {Phys. Rev. Appl.}\ }\textbf {\bibinfo
  {volume} {13}},\ \bibinfo {pages} {034054} (\bibinfo {year}
  {2020})}\BibitemShut {NoStop}%
\bibitem [{\citenamefont
  {Mostafazadeh}(2002{\natexlab{a}})}]{mostafazadeh2002pseudo}%
  \BibitemOpen
  \bibfield  {author} {\bibinfo {author} {\bibfnamefont {A.}~\bibnamefont
  {Mostafazadeh}},\ }\bibfield  {title} {\bibinfo {title} {Pseudo-{Hermiticity}
  versus {PT} symmetry: the necessary condition for the reality of the spectrum
  of a non-{Hermitian} {Hamiltonian}},\ }\href
  {https://doi.org/10.1063/1.1418246} {\bibfield  {journal} {\bibinfo
  {journal} {J. Math. Phys.}\ }\textbf {\bibinfo {volume} {43}},\ \bibinfo
  {pages} {205} (\bibinfo {year} {2002}{\natexlab{a}})}\BibitemShut {NoStop}%
\bibitem [{\citenamefont
  {Mostafazadeh}(2002{\natexlab{b}})}]{mostafazadeh2002pseudoiii}%
  \BibitemOpen
  \bibfield  {author} {\bibinfo {author} {\bibfnamefont {A.}~\bibnamefont
  {Mostafazadeh}},\ }\bibfield  {title} {\bibinfo {title} {Pseudo-{Hermiticity}
  versus {PT-symmetry III}: Equivalence of pseudo-{Hermiticity} and the
  presence of antilinear symmetries},\ }\href
  {https://doi.org/10.1063/1.1489072} {\bibfield  {journal} {\bibinfo
  {journal} {J. Math. Phys.}\ }\textbf {\bibinfo {volume} {43}},\ \bibinfo
  {pages} {3944} (\bibinfo {year} {2002}{\natexlab{b}})}\BibitemShut {NoStop}%
\bibitem [{\citenamefont {Grigoryan}\ and\ \citenamefont
  {Xia}(2022)}]{grigoryan2022pseudo}%
  \BibitemOpen
  \bibfield  {author} {\bibinfo {author} {\bibfnamefont {V.~L.}\ \bibnamefont
  {Grigoryan}}\ and\ \bibinfo {author} {\bibfnamefont {K.}~\bibnamefont
  {Xia}},\ }\bibfield  {title} {\bibinfo {title} {Pseudo-{Hermitian}
  magnon-polariton system with a three-dimensional exceptional surface},\
  }\href {https://doi.org/10.1103/PhysRevB.106.014404} {\bibfield  {journal}
  {\bibinfo  {journal} {Phys. Rev. B}\ }\textbf {\bibinfo {volume} {106}},\
  \bibinfo {pages} {014404} (\bibinfo {year} {2022})}\BibitemShut {NoStop}%
\bibitem [{\citenamefont {Haus}\ and\ \citenamefont
  {Huang}(1991)}]{haus1991coupled}%
  \BibitemOpen
  \bibfield  {author} {\bibinfo {author} {\bibfnamefont {H.~A.}\ \bibnamefont
  {Haus}}\ and\ \bibinfo {author} {\bibfnamefont {W.}~\bibnamefont {Huang}},\
  }\bibfield  {title} {\bibinfo {title} {Coupled-mode theory},\ }\href
  {https://doi.org/10.1109/5.104225} {\bibfield  {journal} {\bibinfo  {journal}
  {Proc. IEEE}\ }\textbf {\bibinfo {volume} {79}},\ \bibinfo {pages} {1505}
  (\bibinfo {year} {1991})}\BibitemShut {NoStop}%
\bibitem [{\citenamefont {Xiong}\ \emph {et~al.}(2021)\citenamefont {Xiong},
  \citenamefont {Li}, \citenamefont {Song}, \citenamefont {Chen}, \citenamefont
  {Zhang},\ and\ \citenamefont {Wang}}]{xiong2021higher}%
  \BibitemOpen
  \bibfield  {author} {\bibinfo {author} {\bibfnamefont {W.}~\bibnamefont
  {Xiong}}, \bibinfo {author} {\bibfnamefont {Z.}~\bibnamefont {Li}}, \bibinfo
  {author} {\bibfnamefont {Y.}~\bibnamefont {Song}}, \bibinfo {author}
  {\bibfnamefont {J.}~\bibnamefont {Chen}}, \bibinfo {author} {\bibfnamefont
  {G.-Q.}\ \bibnamefont {Zhang}},\ and\ \bibinfo {author} {\bibfnamefont
  {M.}~\bibnamefont {Wang}},\ }\bibfield  {title} {\bibinfo {title}
  {Higher-order exceptional point in a pseudo-{Hermitian} cavity optomechanical
  system},\ }\href {https://doi.org/10.1103/PhysRevA.104.063508} {\bibfield
  {journal} {\bibinfo  {journal} {Phys. Rev. A}\ }\textbf {\bibinfo {volume}
  {104}},\ \bibinfo {pages} {063508} (\bibinfo {year} {2021})}\BibitemShut
  {NoStop}%
\bibitem [{\citenamefont {Li}\ \emph {et~al.}(2022)\citenamefont {Li},
  \citenamefont {Li},\ and\ \citenamefont {Zhong}}]{li2022high}%
  \BibitemOpen
  \bibfield  {author} {\bibinfo {author} {\bibfnamefont {Z.}~\bibnamefont
  {Li}}, \bibinfo {author} {\bibfnamefont {X.}~\bibnamefont {Li}},\ and\
  \bibinfo {author} {\bibfnamefont {X.}~\bibnamefont {Zhong}},\ }\bibfield
  {title} {\bibinfo {title} {High-order exceptional point in a nanofiber cavity
  quantum electrodynamics system},\ }\bibfield  {journal} {\bibinfo  {journal}
  {arXiv:2201.03768}\ }\href {https://doi.org/10.48550/ARXIV.2201.03768}
  {10.48550/ARXIV.2201.03768} (\bibinfo {year} {2022})\BibitemShut {NoStop}%
\bibitem [{\citenamefont {Sachs}(1987)}]{sachs1987physics}%
  \BibitemOpen
  \bibfield  {author} {\bibinfo {author} {\bibfnamefont {R.~G.}\ \bibnamefont
  {Sachs}},\ }\href@noop {} {\emph {\bibinfo {title} {The physics of time
  reversal}}}\ (\bibinfo  {publisher} {University of Chicago Press},\ \bibinfo
  {year} {1987})\BibitemShut {NoStop}%
\bibitem [{\citenamefont {Choi}\ \emph {et~al.}(2018)\citenamefont {Choi},
  \citenamefont {Hahn}, \citenamefont {Yoon},\ and\ \citenamefont
  {Song}}]{choi2018observation}%
  \BibitemOpen
  \bibfield  {author} {\bibinfo {author} {\bibfnamefont {Y.}~\bibnamefont
  {Choi}}, \bibinfo {author} {\bibfnamefont {C.}~\bibnamefont {Hahn}}, \bibinfo
  {author} {\bibfnamefont {J.~W.}\ \bibnamefont {Yoon}},\ and\ \bibinfo
  {author} {\bibfnamefont {S.~H.}\ \bibnamefont {Song}},\ }\bibfield  {title}
  {\bibinfo {title} {Observation of an anti-{PT}-symmetric exceptional point
  and energy-difference conserving dynamics in electrical circuit resonators},\
  }\href {https://doi.org/10.1038/s41467-018-04690-y} {\bibfield  {journal}
  {\bibinfo  {journal} {Nat. Commun.}\ }\textbf {\bibinfo {volume} {9}},\
  \bibinfo {pages} {1} (\bibinfo {year} {2018})}\BibitemShut {NoStop}%
\bibitem [{\citenamefont {Luo}\ \emph {et~al.}(2022)\citenamefont {Luo},
  \citenamefont {Zhang},\ and\ \citenamefont {Du}}]{luo2022quantum}%
  \BibitemOpen
  \bibfield  {author} {\bibinfo {author} {\bibfnamefont {X.-W.}\ \bibnamefont
  {Luo}}, \bibinfo {author} {\bibfnamefont {C.}~\bibnamefont {Zhang}},\ and\
  \bibinfo {author} {\bibfnamefont {S.}~\bibnamefont {Du}},\ }\bibfield
  {title} {\bibinfo {title} {Quantum squeezing and sensing with
  pseudo-anti-parity-time symmetry},\ }\href
  {https://doi.org/10.1103/PhysRevLett.128.173602} {\bibfield  {journal}
  {\bibinfo  {journal} {Phys. Rev. Lett.}\ }\textbf {\bibinfo {volume} {128}},\
  \bibinfo {pages} {173602} (\bibinfo {year} {2022})}\BibitemShut {NoStop}%
\bibitem [{\citenamefont {Kim}\ \emph {et~al.}(2022{\natexlab{a}})\citenamefont
  {Kim}, \citenamefont {Yang}, \citenamefont {Gui},\ and\ \citenamefont
  {Hu}}]{kim2022visualization}%
  \BibitemOpen
  \bibfield  {author} {\bibinfo {author} {\bibfnamefont {M.}~\bibnamefont
  {Kim}}, \bibinfo {author} {\bibfnamefont {Y.}~\bibnamefont {Yang}}, \bibinfo
  {author} {\bibfnamefont {Y.}~\bibnamefont {Gui}},\ and\ \bibinfo {author}
  {\bibfnamefont {C.-M.}\ \bibnamefont {Hu}},\ }\bibfield  {title} {\bibinfo
  {title} {Visualization of synchronization zone on the {Bloch} sphere through
  an anti-{PT}-symmetric electrical circuit},\ }\href
  {https://doi.org/10.1063/5.0081693} {\bibfield  {journal} {\bibinfo
  {journal} {AIP Adv.}\ }\textbf {\bibinfo {volume} {12}},\ \bibinfo {pages}
  {035217} (\bibinfo {year} {2022}{\natexlab{a}})}\BibitemShut {NoStop}%
\bibitem [{\citenamefont {Arwas}\ \emph {et~al.}(2022)\citenamefont {Arwas},
  \citenamefont {Gadasi}, \citenamefont {Gershenzon}, \citenamefont {Friesem},
  \citenamefont {Davidson},\ and\ \citenamefont {Raz}}]{arwas2022anyonic}%
  \BibitemOpen
  \bibfield  {author} {\bibinfo {author} {\bibfnamefont {G.}~\bibnamefont
  {Arwas}}, \bibinfo {author} {\bibfnamefont {S.}~\bibnamefont {Gadasi}},
  \bibinfo {author} {\bibfnamefont {I.}~\bibnamefont {Gershenzon}}, \bibinfo
  {author} {\bibfnamefont {A.}~\bibnamefont {Friesem}}, \bibinfo {author}
  {\bibfnamefont {N.}~\bibnamefont {Davidson}},\ and\ \bibinfo {author}
  {\bibfnamefont {O.}~\bibnamefont {Raz}},\ }\bibfield  {title} {\bibinfo
  {title} {Anyonic-parity-time symmetry in complex-coupled lasers},\ }\href
  {https://doi.org/10.1126/sciadv.abm7454} {\bibfield  {journal} {\bibinfo
  {journal} {Sci. Adv.}\ }\textbf {\bibinfo {volume} {8}},\ \bibinfo {pages}
  {eabm7454} (\bibinfo {year} {2022})}\BibitemShut {NoStop}%
\bibitem [{\citenamefont {Yin}\ \emph {et~al.}(2022)\citenamefont {Yin},
  \citenamefont {Huang}, \citenamefont {Ma}, \citenamefont {Hao}, \citenamefont
  {Gao}, \citenamefont {Ma},\ and\ \citenamefont {Dong}}]{yin2022wireless}%
  \BibitemOpen
  \bibfield  {author} {\bibinfo {author} {\bibfnamefont {K.}~\bibnamefont
  {Yin}}, \bibinfo {author} {\bibfnamefont {Y.}~\bibnamefont {Huang}}, \bibinfo
  {author} {\bibfnamefont {C.}~\bibnamefont {Ma}}, \bibinfo {author}
  {\bibfnamefont {X.}~\bibnamefont {Hao}}, \bibinfo {author} {\bibfnamefont
  {X.}~\bibnamefont {Gao}}, \bibinfo {author} {\bibfnamefont {X.}~\bibnamefont
  {Ma}},\ and\ \bibinfo {author} {\bibfnamefont {T.}~\bibnamefont {Dong}},\
  }\bibfield  {title} {\bibinfo {title} {Wireless real-time capacitance readout
  based on perturbed nonlinear parity-time symmetry},\ }\href
  {https://doi.org/10.1063/5.0093982} {\bibfield  {journal} {\bibinfo
  {journal} {Appl. Phys. Lett.}\ }\textbf {\bibinfo {volume} {120}},\ \bibinfo
  {pages} {194101} (\bibinfo {year} {2022})}\BibitemShut {NoStop}%
\bibitem [{\citenamefont {Yang}\ \emph {et~al.}(2021)\citenamefont {Yang},
  \citenamefont {Ye}, \citenamefont {Farhat},\ and\ \citenamefont
  {Chen}}]{yang2021ultrarobust}%
  \BibitemOpen
  \bibfield  {author} {\bibinfo {author} {\bibfnamefont {M.}~\bibnamefont
  {Yang}}, \bibinfo {author} {\bibfnamefont {Z.}~\bibnamefont {Ye}}, \bibinfo
  {author} {\bibfnamefont {M.}~\bibnamefont {Farhat}},\ and\ \bibinfo {author}
  {\bibfnamefont {P.-Y.}\ \bibnamefont {Chen}},\ }\bibfield  {title} {\bibinfo
  {title} {Ultrarobust wireless interrogation for sensors and transducers: A
  non-{Hermitian} telemetry technique},\ }\href
  {https://doi.org/10.1109/TIM.2021.3107057} {\bibfield  {journal} {\bibinfo
  {journal} {IEEE Trans. Instrum. Meas.}\ }\textbf {\bibinfo {volume} {70}},\
  \bibinfo {pages} {1} (\bibinfo {year} {2021})}\BibitemShut {NoStop}%
\bibitem [{\citenamefont {Seo}(2019)}]{seo2019comparative}%
  \BibitemOpen
  \bibfield  {author} {\bibinfo {author} {\bibfnamefont {D.-W.}\ \bibnamefont
  {Seo}},\ }\bibfield  {title} {\bibinfo {title} {Comparative analysis of
  two-and three-coil {WPT} systems based on transmission efficiency},\ }\href
  {https://doi.org/10.1109/ACCESS.2019.2947093} {\bibfield  {journal} {\bibinfo
   {journal} {IEEE Access}\ }\textbf {\bibinfo {volume} {7}},\ \bibinfo {pages}
  {151962} (\bibinfo {year} {2019})}\BibitemShut {NoStop}%
\bibitem [{sup()}]{supplementary}%
  \BibitemOpen
  \href@noop {} {}\bibinfo {note} {See Supplement Material at \emph{url} for
  implementation of gain elements and detailed information on the simulation
  and experimental setup.}\BibitemShut {Stop}%
\bibitem [{\citenamefont {Ye}\ \emph {et~al.}(2020)\citenamefont {Ye},
  \citenamefont {Farhat},\ and\ \citenamefont {Chen}}]{ye2020tunability}%
  \BibitemOpen
  \bibfield  {author} {\bibinfo {author} {\bibfnamefont {Z.}~\bibnamefont
  {Ye}}, \bibinfo {author} {\bibfnamefont {M.}~\bibnamefont {Farhat}},\ and\
  \bibinfo {author} {\bibfnamefont {P.-Y.}\ \bibnamefont {Chen}},\ }\bibfield
  {title} {\bibinfo {title} {Tunability and switching of {Fano and Lorentz}
  resonances in {PTX}-symmetric electronic systems},\ }\href
  {https://doi.org/10.1063/5.0014919} {\bibfield  {journal} {\bibinfo
  {journal} {Appl. Phys. Lett.}\ }\textbf {\bibinfo {volume} {117}},\ \bibinfo
  {pages} {031101} (\bibinfo {year} {2020})}\BibitemShut {NoStop}%
\bibitem [{\citenamefont {Kim}\ \emph {et~al.}(2022{\natexlab{b}})\citenamefont
  {Kim}, \citenamefont {Yoo}, \citenamefont {Joo}, \citenamefont {Lee},
  \citenamefont {An}, \citenamefont {Nam}, \citenamefont {Han}, \citenamefont
  {Kim},\ and\ \citenamefont {Kim}}]{kim2022wide}%
  \BibitemOpen
  \bibfield  {author} {\bibinfo {author} {\bibfnamefont {H.}~\bibnamefont
  {Kim}}, \bibinfo {author} {\bibfnamefont {S.}~\bibnamefont {Yoo}}, \bibinfo
  {author} {\bibfnamefont {H.}~\bibnamefont {Joo}}, \bibinfo {author}
  {\bibfnamefont {J.}~\bibnamefont {Lee}}, \bibinfo {author} {\bibfnamefont
  {D.}~\bibnamefont {An}}, \bibinfo {author} {\bibfnamefont {S.}~\bibnamefont
  {Nam}}, \bibinfo {author} {\bibfnamefont {H.}~\bibnamefont {Han}}, \bibinfo
  {author} {\bibfnamefont {D.-H.}\ \bibnamefont {Kim}},\ and\ \bibinfo {author}
  {\bibfnamefont {S.}~\bibnamefont {Kim}},\ }\bibfield  {title} {\bibinfo
  {title} {Wide-range robust wireless power transfer using heterogeneously
  coupled and flippable neutrals in parity-time symmetry},\ }\href
  {https://doi.org/10.1126/sciadv.abo4610} {\bibfield  {journal} {\bibinfo
  {journal} {Sci. Adv.}\ }\textbf {\bibinfo {volume} {8}},\ \bibinfo {pages}
  {eabo4610} (\bibinfo {year} {2022}{\natexlab{b}})}\BibitemShut {NoStop}%
\bibitem [{\citenamefont {Qu}\ \emph {et~al.}(2022)\citenamefont {Qu},
  \citenamefont {Zhang}, \citenamefont {Gu},\ and\ \citenamefont
  {Shu}}]{qu2022wireless}%
  \BibitemOpen
  \bibfield  {author} {\bibinfo {author} {\bibfnamefont {Y.}~\bibnamefont
  {Qu}}, \bibinfo {author} {\bibfnamefont {B.}~\bibnamefont {Zhang}}, \bibinfo
  {author} {\bibfnamefont {W.}~\bibnamefont {Gu}},\ and\ \bibinfo {author}
  {\bibfnamefont {X.}~\bibnamefont {Shu}},\ }\bibfield  {title} {\bibinfo
  {title} {Wireless power transfer system with high-order compensation network
  based on parity-time-symmetric principle and relay coil},\ }\bibfield
  {journal} {\bibinfo  {journal} {IEEE Trans. Power Electron.}\ }\href
  {https://doi.org/10.1109/TPEL.2022.3201693} {10.1109/TPEL.2022.3201693}
  (\bibinfo {year} {2022})\BibitemShut {NoStop}%
\bibitem [{\citenamefont {Mohseni}\ \emph {et~al.}(2022)\citenamefont
  {Mohseni}, \citenamefont {Nikzamir}, \citenamefont {Cao},\ and\ \citenamefont
  {Capolino}}]{mohseni2022one}%
  \BibitemOpen
  \bibfield  {author} {\bibinfo {author} {\bibfnamefont {F.}~\bibnamefont
  {Mohseni}}, \bibinfo {author} {\bibfnamefont {A.}~\bibnamefont {Nikzamir}},
  \bibinfo {author} {\bibfnamefont {H.}~\bibnamefont {Cao}},\ and\ \bibinfo
  {author} {\bibfnamefont {F.}~\bibnamefont {Capolino}},\ }\bibfield  {title}
  {\bibinfo {title} {One-transmitter multiple-receiver wireless power transfer
  system using an exceptional point of degeneracy},\ }\bibfield  {journal}
  {\bibinfo  {journal} {arXiv:2204.10928}\ }\href
  {https://doi.org/10.48550/arXiv.2204.10928} {10.48550/arXiv.2204.10928}
  (\bibinfo {year} {2022})\BibitemShut {NoStop}%
\end{thebibliography}%


\begin{thebibliography}{2}%
\makeatletter
\providecommand \@ifxundefined [1]{%
 \@ifx{#1\undefined}
}%
\providecommand \@ifnum [1]{%
 \ifnum #1\expandafter \@firstoftwo
 \else \expandafter \@secondoftwo
 \fi
}%
\providecommand \@ifx [1]{%
 \ifx #1\expandafter \@firstoftwo
 \else \expandafter \@secondoftwo
 \fi
}%
\providecommand \natexlab [1]{#1}%
\providecommand \enquote  [1]{``#1''}%
\providecommand \bibnamefont  [1]{#1}%
\providecommand \bibfnamefont [1]{#1}%
\providecommand \citenamefont [1]{#1}%
\providecommand \href@noop [0]{\@secondoftwo}%
\providecommand \href [0]{\begingroup \@sanitize@url \@href}%
\providecommand \@href[1]{\@@startlink{#1}\@@href}%
\providecommand \@@href[1]{\endgroup#1\@@endlink}%
\providecommand \@sanitize@url [0]{\catcode `\\12\catcode `\$12\catcode
  `\&12\catcode `\#12\catcode `\^12\catcode `\_12\catcode `\%12\relax}%
\providecommand \@@startlink[1]{}%
\providecommand \@@endlink[0]{}%
\providecommand \url  [0]{\begingroup\@sanitize@url \@url }%
\providecommand \@url [1]{\endgroup\@href {#1}{\urlprefix }}%
\providecommand \urlprefix  [0]{URL }%
\providecommand \Eprint [0]{\href }%
\providecommand \doibase [0]{https://doi.org/}%
\providecommand \selectlanguage [0]{\@gobble}%
\providecommand \bibinfo  [0]{\@secondoftwo}%
\providecommand \bibfield  [0]{\@secondoftwo}%
\providecommand \translation [1]{[#1]}%
\providecommand \BibitemOpen [0]{}%
\providecommand \bibitemStop [0]{}%
\providecommand \bibitemNoStop [0]{.\EOS\space}%
\providecommand \EOS [0]{\spacefactor3000\relax}%
\providecommand \BibitemShut  [1]{\csname bibitem#1\endcsname}%
\let\auto@bib@innerbib\@empty
\bibitem [{\citenamefont {Assawaworrarit}\ \emph {et~al.}(2017)\citenamefont
  {Assawaworrarit}, \citenamefont {Yu},\ and\ \citenamefont
  {Fan}}]{assawaworrarit2017robust}%
  \BibitemOpen
  \bibfield  {author} {\bibinfo {author} {\bibfnamefont {S.}~\bibnamefont
  {Assawaworrarit}}, \bibinfo {author} {\bibfnamefont {X.}~\bibnamefont {Yu}},\
  and\ \bibinfo {author} {\bibfnamefont {S.}~\bibnamefont {Fan}},\ }\bibfield
  {title} {\bibinfo {title} {Robust wireless power transfer using a nonlinear
  parity-time-symmetric circuit},\ }\href {https://doi.org/10.1038/nature22404}
  {\bibfield  {journal} {\bibinfo  {journal} {Nature}\ }\textbf {\bibinfo
  {volume} {546}},\ \bibinfo {pages} {387} (\bibinfo {year}
  {2017})}\BibitemShut {NoStop}%
\bibitem [{\citenamefont {Yin}\ \emph {et~al.}(2022)\citenamefont {Yin},
  \citenamefont {Huang}, \citenamefont {Ma}, \citenamefont {Hao}, \citenamefont
  {Gao}, \citenamefont {Ma},\ and\ \citenamefont {Dong}}]{yin2022wireless}%
  \BibitemOpen
  \bibfield  {author} {\bibinfo {author} {\bibfnamefont {K.}~\bibnamefont
  {Yin}}, \bibinfo {author} {\bibfnamefont {Y.}~\bibnamefont {Huang}}, \bibinfo
  {author} {\bibfnamefont {C.}~\bibnamefont {Ma}}, \bibinfo {author}
  {\bibfnamefont {X.}~\bibnamefont {Hao}}, \bibinfo {author} {\bibfnamefont
  {X.}~\bibnamefont {Gao}}, \bibinfo {author} {\bibfnamefont {X.}~\bibnamefont
  {Ma}},\ and\ \bibinfo {author} {\bibfnamefont {T.}~\bibnamefont {Dong}},\
  }\bibfield  {title} {\bibinfo {title} {Wireless real-time capacitance readout
  based on perturbed nonlinear parity-time symmetry},\ }\href
  {https://doi.org/10.1063/5.0093982} {\bibfield  {journal} {\bibinfo
  {journal} {Appl. Phys. Lett.}\ }\textbf {\bibinfo {volume} {120}},\ \bibinfo
  {pages} {194101} (\bibinfo {year} {2022})}\BibitemShut {NoStop}%
\end{thebibliography}%

\end{document}


\title{Supplemental material: Frequency-stable robust wireless power transfer \\ based on high-order pseudo-Hermitian physics}

\author{Xianglin Hao}
\orcid{0000-0001-7149-0113}
\thanks{These authors contributed equally to this work.}

\author{Ke Yin}
\orcid{0000-0002-8534-216X}
\thanks{These authors contributed equally to this work.}

\author{Jianlong Zou}
\orcid{0000-0003-1489-0828}
\email{superzou@mail.xjtu.edu.cn}

\author{Ruibin Wang}

\author{Yuangen Huang}
\orcid{0000-0002-2486-2778}

\author{Xikui Ma}

\author{Tianyu Dong}
\orcid{0000-0003-4816-0073}
\email[Author to whom correspondence should be addressed. Please e-mail to: ]{tydong@mail.xjtu.edu.cn}
\affiliation{School of Electrical Engineering, Xi’an Jiaotong University, Xi’an 710049, China}%

\date{October 11, 2022}

\begin{abstract}
In the supplementary, the implementation of the nonlinear gain elements, power transmission efficiency and the experiment setup are discussed.
\end{abstract}

\maketitle

\section{Implementation of nonlinear gain elements}
\begin{figure}[!ht]
    \centering
    \includegraphics[width=3.3in]{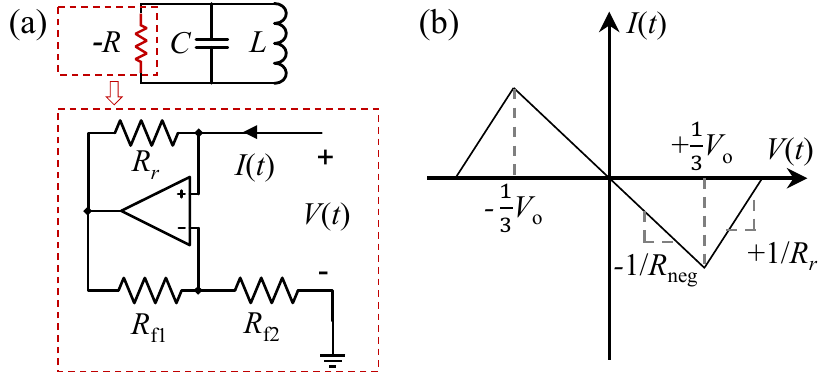}
    \caption{(a) Nonlinear gain element and (b) its piecewise linear $I$-$V$ relationship. $R_\text{f1}=2R_\text{f2}$ in our letter. $V_o$ is the maximum output voltage of the operational amplifier and $R_\text{neg}=R_r/2$.}
    \label{fig:figS1}
\end{figure}

In our work, the nonlinear gain element consists of an operational amplifier (op-amp) and resistors, which has been extensively studied in previous works \cite{assawaworrarit2017robust,yin2022wireless}. \figref{fig:figS1}(a) shows the circuit diagram of the gain element. 
For such a non-inverting amplifier, due to the resistance $R_r$ between the non-inverting input port and the output port of the op-amp, the whole circuit behaves like a resistor with negative resistance in the linear output range of the op-amp, \emph{i.e.}, $-R = V(t) / I(t)$. The actual $I-V$ relation of such a gain element is a piece-wise linear function, as depicted in \figref{fig:figS1}(b).And the current with respect to the voltage amplitude can be expressed as
\begin{equation}
    I(V)=
    \begin{cases}
   -\frac{V}{R_\text{neg}}, & |V|\leq V_o/3 \\[0.5em]
   \frac{V-V_o}{R_r}, & |V|>V_o/3
   \end{cases}.
\end{equation}
where $R_r$ is the resistance shown in \figref{fig:figS1}(a), $R_\text{neg} = R_r/2$ and $V_o$ is the maximum output voltage of the op-amp.

The voltage-dependent gain reads as $g(|V|) = R^{-1} \sqrt{L/C} = \sqrt{L/C} \text{d}I / \text{d}V$ \cite{yin2022wireless}. When the op-amp is not saturated, the non-inverting amplifier circuit is equivalent to a fixed negative resistance $-R_\text{neq}$; and initial unsaturated gain $g_i=\sqrt{L/C}/R_\text{neq}$. For the saturation state when $|V|>V_o/3$ ($V_o = 5$~\si{V} in our work), the equivalent resistance of the amplifier circuit reads $R_r$; and saturated gain $g_s = -\sqrt{L/C}/R_o$. Given that $V_1$ be the voltage amplitude of $V(t) = V_1 \sin (\omega t)$, the equivalent gain strength in one cycle reads  $g = (2\pi)^{-1} \int g(|V|) \text{d}V$, \emph{i.e.},
\begin{equation}
    g =
    \begin{cases}
       g_i, & V_1 \leq V_o/3 \\[0.5em]
       \frac{2}{\pi} (g_i-g_s) \sin^{-1}\left( \frac{V_o}{3V_1} \right) + g_s, & V_1 > V_o/3
    \end{cases}.
\end{equation}
It is evident that the gain strength $g(|V|)$ is decreased as the voltage is increased when the gain element is saturated, with the maximum gain being the unsaturated gain $g_i$. Therefore, one can just let the unsaturated gain be greater than the maximum theoretically required gain in the whole coupling region. As the saturation degree deepens (\emph{i.e.}, $V_1$ is increased), $g$ decreases gradually until the system goes into a steady state.

\section{Power transmission efficiency}
\begin{figure}[!ht]
    \centering
    \includegraphics[width=3.3in]{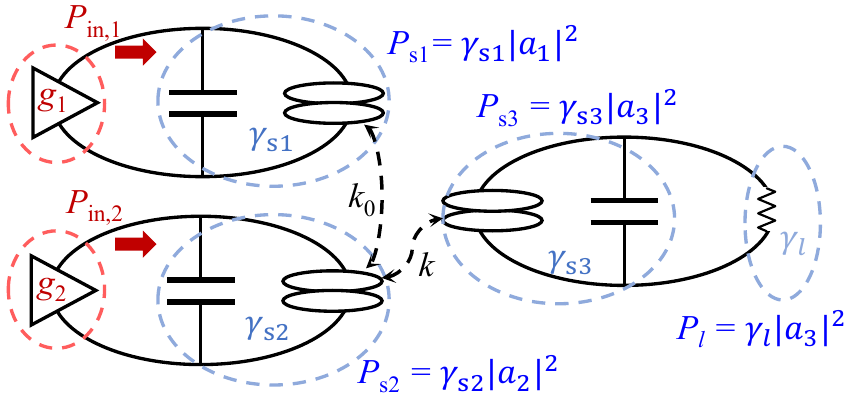}
    \caption{(color online) Schematic of the power flow in the considered three-coil WPT system. Red and blue dotted ellipse regions represent power sources and power loss, respectively. The loss coefficients of the three resonators are $\gamma_{\mathrm{s1}}$, $\gamma_\text{s2}$, $\gamma_\text{s3}$, respectively. The loss coefficient of the load resistor is $\gamma_l$.}
    \label{fig:figS2}
\end{figure}

The efficiency can be directly obtained according to the mode amplitudes without calculating the actual power. \figref{fig:figS2} shows the power flow of the proposed system. 
Both two gain elements inject energy into the resonator. Some of the input power is transferred to the load resistor, and the rest is dissipated in the LC resonators since the resonators have internal resistance. The input power is equal to the total power loss consumed by resonators and load, \emph{i.e.}, $P_{\text{in},1} + P_{\text{in},2} = P_\text{s1} + P_\text{s2} + P_\text{s3} + P_l$. According to the coupled mode theory, the total power transmitted to the receiving coil is calculated by $P_\text{s3} + P_l = \gamma |a_3|^2$ where $\gamma=\gamma_l + \gamma_\text{s3}$; the power consumed by the load reads $P_l = \gamma_l |a_3|^2$; and the power losses at the transmitter end read $P_{s,n} = \gamma_{s,n} |a_n|^2$ ($n=1,2$). As a result, the steady-state power transfer efficiency, defined as the ratio between the power delivered to the load and the power input to the two transmit coils, can be expressed as
\begin{equation}\label{eq:s3}
\begin{split}
    \eta_\text{PTE} &= \frac{P_l}{P_\text{s1} + P_\text{s2} + P_\text{s3} + P_l} \\ 
    &= \frac{\gamma_l \left|a_3\right|^2}{\gamma_\text{s1}\left|a_1\right|^2+\gamma_\text{s2}\left|a_2\right|^2+\gamma\left|a_3\right|^2} .
\end{split}
\end{equation}

\section{Experimental setup}
\begin{figure}[!ht]
    \centering
    \includegraphics[width=3.3in]{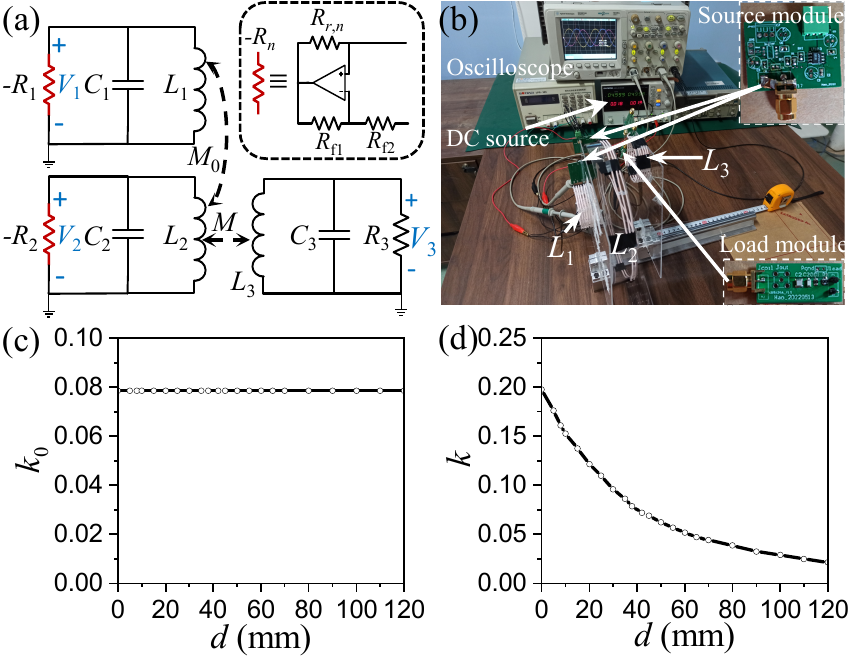}
    \caption{(color online) (a) Circuit schematic and (b) experimental prototype of the third-order pseudo-Hermitian wireless power transfer system. The inset in (a) illustrates the realization of negative resistance $-R$ by utilizing op-amp. Measured results (hollow markers) of the coupling coefficients (c) $k_0$ and (d) $k$ with respect to the distance $d$ between the transmitter coil II $L_2$ and the receiver $L_3$. The solid black line in (c) and (d) are the fitted curves to the measured data points. The relative positions of the coils $L_1$ and $L_2$ remain unchanged in the experiments.}
    \label{fig:figS3}
\end{figure}

\figref{fig:figS3}(a) shows the circuit diagram, which was simulated by the commercial circuit simulation software LTspice. Here, the op-amp, configured as a non-inverting amplifier, together with an resistor with resistance $R_{r,n}$ ($n=1, 2$ for the transmitter I and II, respectively), functions as a nonlinear gain element in each transmitter whose gain coefficients $g_n=\sqrt{L_1/C_1}/R_n$ are in the linear range of the op-amp, where $R_n=(R_{r,n} R_\text{f2})/R_\text{f1}$. The component parameters used in the simulation are summarized in \tabref{tab:table1}. 
\begin{table}[!ht]
    \caption{Component parameters for simulations and experiments}\label{tab:table1}
    \begin{tabular}{cccc}
    \hline
    Parameters & Transmitter I & Transmitter II & Receiver\\
            \hline  
            $L$~[\si{\micro\henry}]  & 3.16 & 3.15 & 3.16 \\
            $C$~[\si{nF}] & 2.01 & 2.00 & 2.01 \\
            $R_{r,n}$~[\si{\kilo\ohm}] & 0.7 & 15.6 & -- \\
            $R_\text{f1}$~[\si{\kilo\ohm}] & 2 & 2 & -- \\
            $R_\text{f2}$~[\si{\kilo\ohm}] & 1 & 1 & -- \\
            $R_3$~[\si{\ohm}] & -- & -- & 500\\
            \hline 
    \end{tabular}
\end{table}
The resistances $R_{r,n}$ are obtained by fine-tuning the resistance value corresponding to $k=0.8$ in consideration of the actual coil loss. In the simulation, all other circuit parameters, \emph{e.g.}, $k_0$, remain unchanged as the coupling coefficient $k$ varies. In addition to the circuit simulation, we constructed an experimental prototype of the pseudo-Hermitian WPT system that constitutes of dual transmitters, as shown in \figref{fig:figS3}(b). The resonant coils used in the experiments are all wound with 0.05~\si{mm}$\times$100 Litz wire. To eliminate the mutual coupling between the transmitter I $L_1$ and the receiver $L_3$ as much as possible, we placed the two coils eccentrically. The receiver resonator is mounted on a linear guide and can be adjusted axially without radial offsets. All the three resonant coils are tuned to have the same resonant frequency of 2~\si{MHz}. The coils $L_1$ and $L_3$ have a measured intrinsic quality factor of $Q_1 = Q_3 = 400$, while it reads $Q_2=286$ for the transmitter II $L_2$, \emph{i.e.}, $\gamma_\text{s1}=\gamma_\text{s3}=1/400$ and $\gamma_\text{s2}=1/286$. \figref{fig:figS3}(c) and \figref{fig:figS3}(d) plot the measured coupling coefficients $k_0$ and $k$ as functions of the separation distance $d$ between the transmitter II $L_2$ and  the receiver $L_3$, respectively. It is evident that $k_0$ is keeping about 0.078 when the distance $d$ varies, implicating that the coupling between $L_1$ and $L_3$ is almost negligible.

\bibliography{supp} 